\documentclass{IEEEoj}

\usepackage{cite}
\usepackage{amsmath,amssymb,amsfonts}
\usepackage{algorithmic}
\usepackage{graphicx,color}
\usepackage{textcomp}
\usepackage{amsthm}
\usepackage{braket}
\usepackage{multirow}
\usepackage{threeparttable}
\usepackage[caption=false,font=footnotesize]{subfig}
\usepackage{siunitx}
\newtheorem{remark}{Remark}
\def\BibTeX{{\rm B\kern-.05em{\sc i\kern-.025em b}\kern-.08em
    T\kern-.1667em\lower.7ex\hbox{E}\kern-.125emX}}
\AtBeginDocument{\definecolor{ojcolor}{cmyk}{0.93,0.59,0.15,0.02}}
\def\OJlogo{\vspace{-4pt}\hskip-4pt\includegraphics[height=18pt]{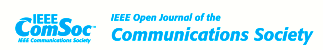}}

\usepackage{hyperref}

\renewcommand{\OJlogo}{}

\begin{document}
\receiveddate{XX Month, XXXX}
\reviseddate{XX Month, XXXX}
\accepteddate{XX Month, XXXX}
\publisheddate{XX Month, XXXX}
\currentdate{11 January, 2024}
\doiinfo{...}

\title{Engineering Quantum Links: Noise and Quantum-State-Degradation Metrics over Metropolitan Fiber Network}

\author{Marcello Caleffi \IEEEmembership{(Senior~Member,~IEEE)}, Laura d'Avossa \IEEEmembership{(Graduate~Student~Member,~IEEE)}, Angela Sara Cacciapuoti
 \IEEEmembership{(Senior~Member,~IEEE)}}
\affil{www.QuantumInternet.it research group, University of Naples Federico II, Naples, 80125 Italy.}
\corresp{CORRESPONDING AUTHOR: Marcello Caleffi (e-mail: marcello.caleffi@unina.it).}
\authornote{This work has been funded by the European Union under Horizon Europe ERC-CoG grant QNattyNet, n.101169850. Views and opinions expressed are however those of the author(s) only and do not necessarily reflect those of the European Union or the European Research Council Executive Agency. Neither the European Union nor the granting authority can be held responsible for them.}
\markboth{Engineering Quantum Links}{Caleffi \textit{et al.}}

\begin{abstract}
Deploying quantum networks over existing network infrastructures requires the same engineering foundations that underpin classical communications: \textit{quantitative models of the channel's noise and of the impairments it imposes on the transmitted information}. In this work, we build such a foundation on experimental measurements, grounding the quantum-network counterparts of the two cornerstone metrics of classical link characterization -- namely, the \textit{SINR} and the \textit{BER} -- on a $\simeq 7.3$~km deployed metropolitan-scale fiber-loop interconnecting two campuses of the University of Naples Federico II within the national \textit{QuantumInternet.it} testbed. On the noise side, we adopt a photon-counting quantum analog of the SINR -- in which dark counts constitute the intrinsic noise and the photons generated by classical traffic (through either spontaneous Raman scattering or inter-fiber crosstalk) constitute the interference -- and we quantify each contribution directly on the deployed loop. On the bit-error side, we consider the main \textit{degrees-of-freedom} available to encode a quantum state within an optical photon -- namely, \textit{polarization}, \textit{time}, and \textit{frequency} -- and we quantify for each degree the channel-induced degradation and its drift over time. These results show that a quantum fiber link, like its classical counterpart, can be captured by a small set of measurable parameters, turning quantum networking over deployed fiber from a physics demonstration into an engineering design problem. Together, they provide the key ingredients of a quantum link budget for the Quantum Internet.
\end{abstract}


\maketitle

\section{INTRODUCTION}
\label{sec:1}

\begin{figure*}
    \centering
    \includegraphics[width=.75\linewidth]{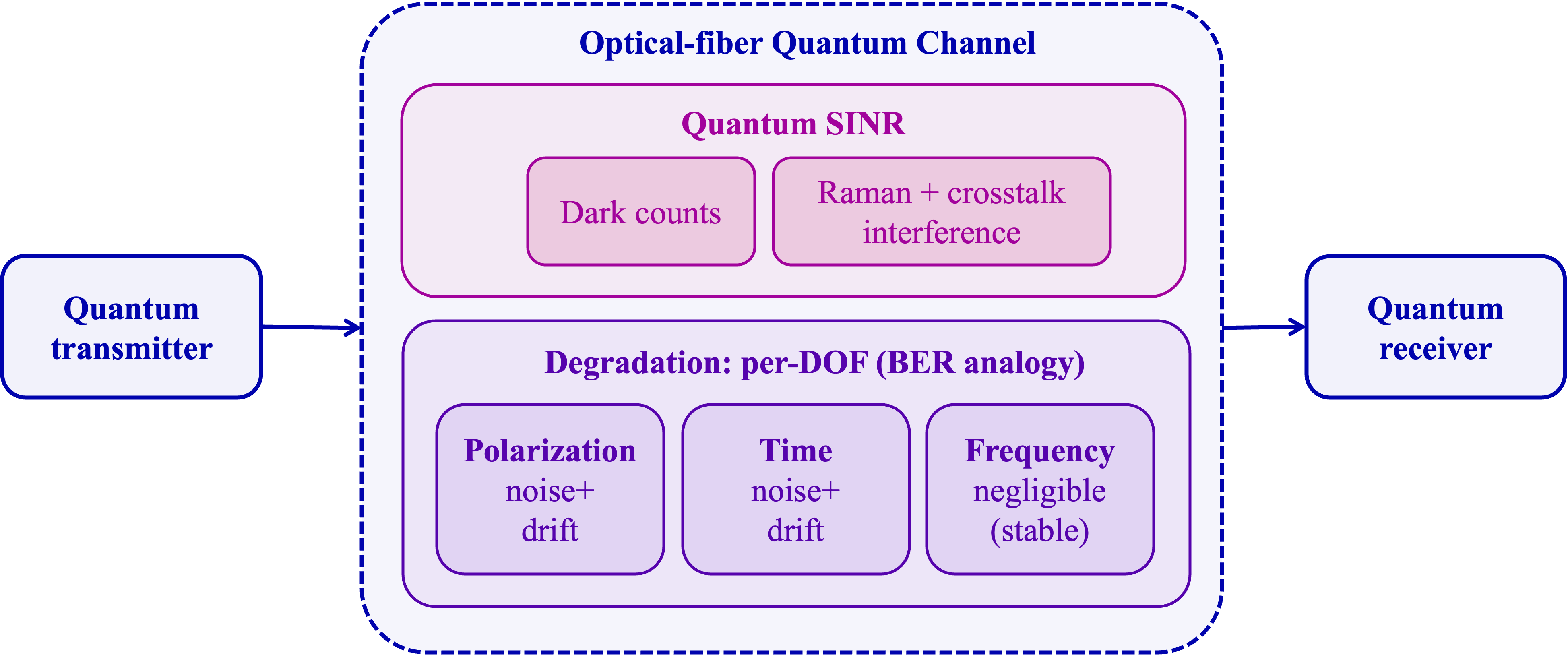}
    \caption{A quantum state propagating through an optical-fiber channel is degraded along a small set of measurable axes, in direct analogy with the impairments characterized in classical links. Two complementary descriptions capture this degradation. On the noise side, the photon-counting quantum-analog of the SINR accounts for the intrinsic noise and the interference introduced by co-propagating classical traffic via spontaneous Raman scattering and inter-fiber crosstalk. On the error side, the analog of the bit-error rate is resolved across the three main degrees-of-freedom available to encode a quantum state onto a single optical photon -- namely, polarization, time, and frequency -- and for each degree we quantify the channel-induced noise and its drift over time. Quantifying these effects provides the building blocks for engineering-oriented quantum-link budgets and metrics.}
    \hrulefill
    \label{fig:new:01}
\end{figure*}

\IEEEPARstart{T}{he} realization of the Quantum Internet promises communication capabilities with no classical counterpart, from quantum teleportation and distributed quantum computing to advanced quantum sensing\cite{CacCalTaf-19, RFC9340}. As quantum hardware matures, the field is now shifting from isolated proof-of-concept demonstrations toward the engineering of scalable quantum networks deployed over real infrastructure \cite{DavMonGri-26, CalDavFlo-26}.

This shift exposes a gap: while the physics of entanglement distribution is increasingly well understood, its engineering description with a network engineering perspective -- namely, the definition of quantitative, reusable \textit{models} that let a network engineer dimension and predict performance of a quantum link -- remains largely absent.

In this work, we address this gap by characterizing optical-fiber quantum links not as physics experiments, but as engineering objects described by a small set of measurable parameters. Crucially, we do so on a deployed metropolitan-scale fiber-loop that has been in operational service for over ten years -- thus inheriting the splices, connectors, and aging that any real-world quantum link must contend with -- rather than on pristine laboratory spools. In doing so, we take a first step toward a link budget for the Quantum Internet: a quantitative account of how deployed fibers degrade quantum information, expressed in the same engineering language that underpins classical communication systems.

\subsection{ANALOGY WITH CLASSICAL CHANNEL MODELS}
\label{sec:1.1}

Our approach is guided by a direct analogy with classical link characterization, which rests on two cornerstone metrics: the \textit{signal-to-interference-plus-noise ratio} (SINR), governing the achievable rate and reliable distance, and the \textit{bit-error rate} (BER) quantifying the corruption of the transmitted information. Together they form the backbone of link-level network engineering across virtually every deployed communication technology. 

Such an analogy, however, is meaningful only once the quantum information carrier is fixed, since the very form of these metrics depends on both i) how quantum information is carried, and ii) the medium through which it propagates. We fix two operative choices.

First, we adopt the \textit{discrete-variable} (DV) paradigm, in which information is encoded onto individual photons and recovered by photon counting, rather than the continuous-variable (CV) paradigm. The metrics we define are consequently expressed in photon counts, whereas their CV analogs would take a different form\footnote{For entanglement distribution coexisting with classical traffic, the DV paradigm is particularly suited: photon counting with narrow coincidence windows suppresses the injected noise by exploiting the temporal correlations of entangled pairs, whereas CV homodyne detection admits such excess noise directly into the measured quadratures.}.

Second, we focus on optical fibers as the transmission medium -- rather than free-space or satellite links -- since at the metropolitan scale considered here it lets quantum networks be built on the already-installed networks, keeping deployment costs low by reusing existing infrastructure rather than requiring dedicated one. Satellite links, by contrast, are better suited to the long-haul and intercontinental backbones where fiber attenuation becomes prohibitive, and lie outside the scope of this work.

Within the fiber, we allocate the quantum plane to the C-band, where single-photon transmission suffers the lowest attenuation (approximately 0.2 dB/km), and relegate classical traffic to the O-band. This allocation -- opposite to that of most prior coexistence studies \cite{TalHesDav-26, TalHesJor-26} -- reserves the lowest-loss portion of the spectrum for the fragile quantum signal (which cannot be amplified), while relegating the robust classical signal to a higher-loss band it can tolerate.

Within this setting, we measure directly analogous metrics for a quantum link, each recast in terms of the physical carriers of optical quantum communications -- namely, single photons rather than optical power -- as shown in Fig.~\ref{fig:new:01}.

On the noise side, we adopt a photon-counting analog of the SINR, in which the signal is the count of detected quantum carriers, the noise is the dark-count rate, and the interference is the count of spurious photons injected by co-propagating traffic. On the error side, the role of the BER is played by the degradation of the encoded quantum state. Indeed, quantum information can be encoded in any of several photonic degrees of freedom -- the main ones being \textit{polarization}, \textit{time}, and \textit{frequency} -- and the channel acts differently on each. We characterize all three, quantifying for every degree both the channel-induced degradation and its drift over time. 

This two-metric framing organizes the remainder of the paper.

\subsection{CONTRIBUTIONS}
\label{sec:1.2}

First, we cast the characterization of an optical-fiber quantum link in the language of classical link engineering, by focusing on a photon-counting analog of the SINR and a per-degree-of-freedom analog of the BER as the two metrics that jointly describe the degradation induced by the channel. To the best of our knowledge, this is the first work to organize the characterization of a deployed-fiber into a unified two-metric framework with a network engineering perspective, spanning noise and all three encoding degrees of freedom, rather than addressing individual impairments in isolation.

Second, we ground this framework on experimental measurements carried out not only in a controlled laboratory, but over a $\simeq 7.3$~km metropolitan fiber-loop that has been in operational service for more than ten years, now part of the national \textit{QuantumInternet.it testbed}. This lets us quantify, on infrastructure that already carries production traffic, the noise and drift that a deployed quantum link must contend with, rather than the idealized behavior of pristine laboratory fiber.

Third, on the noise side, we develop and experimentally validate a predictive model of the C-band Raman noise induced by co-propagating O-band classical traffic. The model reveals a characteristic noise minimum near 1535 nm that the widely used Hollenbeck–Cantrell model \cite{HolCan-02} fails to capture -- deviating from it by up to $19.5 \%$ -- and its accuracy is confirmed across three independent classical sources and multiple fiber lengths. Requiring only experimentally accessible parameters, it directly informs the optimal allocation of quantum channels in the DWDM ITU grid.

Fourth, on the error side, we characterize the channel-induced degradation and its temporal drift across all three encoding degrees of freedom.

Fifth, we distill the degradation results into two compact, closed-form per-link models -- a stochastic-rotation model of polarization drift and a power-law model of temporal drift -- whose few measurable parameters let a designer predict the fidelity, and the recalibration rates, that a deployed fiber used as  quantum link demands. Together with the noise contributions, these provide the first ingredients of a quantum-network link budget.

\subsection{RELATED WORKS}
\label{sec:1.3}

The characterization of fiber-based quantum channels has been approached along two lines, against which we position our contribution.

\textit{Quantum–classical coexistence and Raman noise}. A first line studies the noise that co-propagating classical traffic injects into a shared fiber, dominated by spontaneous Raman scattering. In \cite{ThoKanKum-23}, the authors characterize entangled-pair channels co-propagating with high-power classical light over installed fiber and show that some quantum–classical wavelength combinations markedly outperform others, pointing to an optimal allocation, which they further develop for minimal-noise channel selection. In \cite{ChaLukAls-23} the authors characterize a coexistent quantum channel via spectrally resolved Bayesian process tomography and derive a working model for network design. These works establish the noise side in depth, but each treats it in isolation: a single impairment class, and typically a single encoding.

\textit{Deployed metropolitan testbeds}. A second line distributes entanglement over deployed metropolitan fiber and contends with real-world impairments. Recent testbeds demonstrate entanglement distribution with co-propagating classical traffic and active polarization stabilization over installed fiber \cite{SenFlaAnd-25}, while programmatic efforts such as IEQNET and DC-QNet operate multi-node networks over deployed fiber \cite{ChuKanLau-21}. These works demonstrate and stabilize working links, but do not extract a reusable, metric-based channel description that an engineer could use to predict and dimension performance.

\textit{Positioning}. Our work draws on both lines but differs in scope: rather than treating noise, polarization, timing, and spectral behavior as separate problems, we organize them into a single framework built on two complementary metrics, populated with measurements on a fiber in operational service for over a decade. It is this unification, absent from the works above, that our contributions build on.


\subsection{OUTLINE}
\label{sec:1.4}

The remainder of the paper is organized as follows. Section~\ref{sec:2} describes the experimental setup within the national \textit{QuantumInternet.it testbed}. Section~\ref{sec:3} addresses the noise aspects of the channel modeling, introducing the photon-counting SINR and quantifying its two interference contributions, whereas Section~\ref{sec:4} addresses the error aspects, characterizing the channel-induced degradation and its temporal drift across the three considered encoding degrees of freedom. Finally, Section~\ref{sec:5} assembles these results into the first ingredients of a quantum link budget and Section~\ref{sec:6} discusses the implications and outlines future directions. 

\section{EXPERIMENTAL SETUP OVERVIEW}
\label{sec:2}

\begin{figure*}[t]
    \centering
    \includegraphics[width=0.90\textwidth]{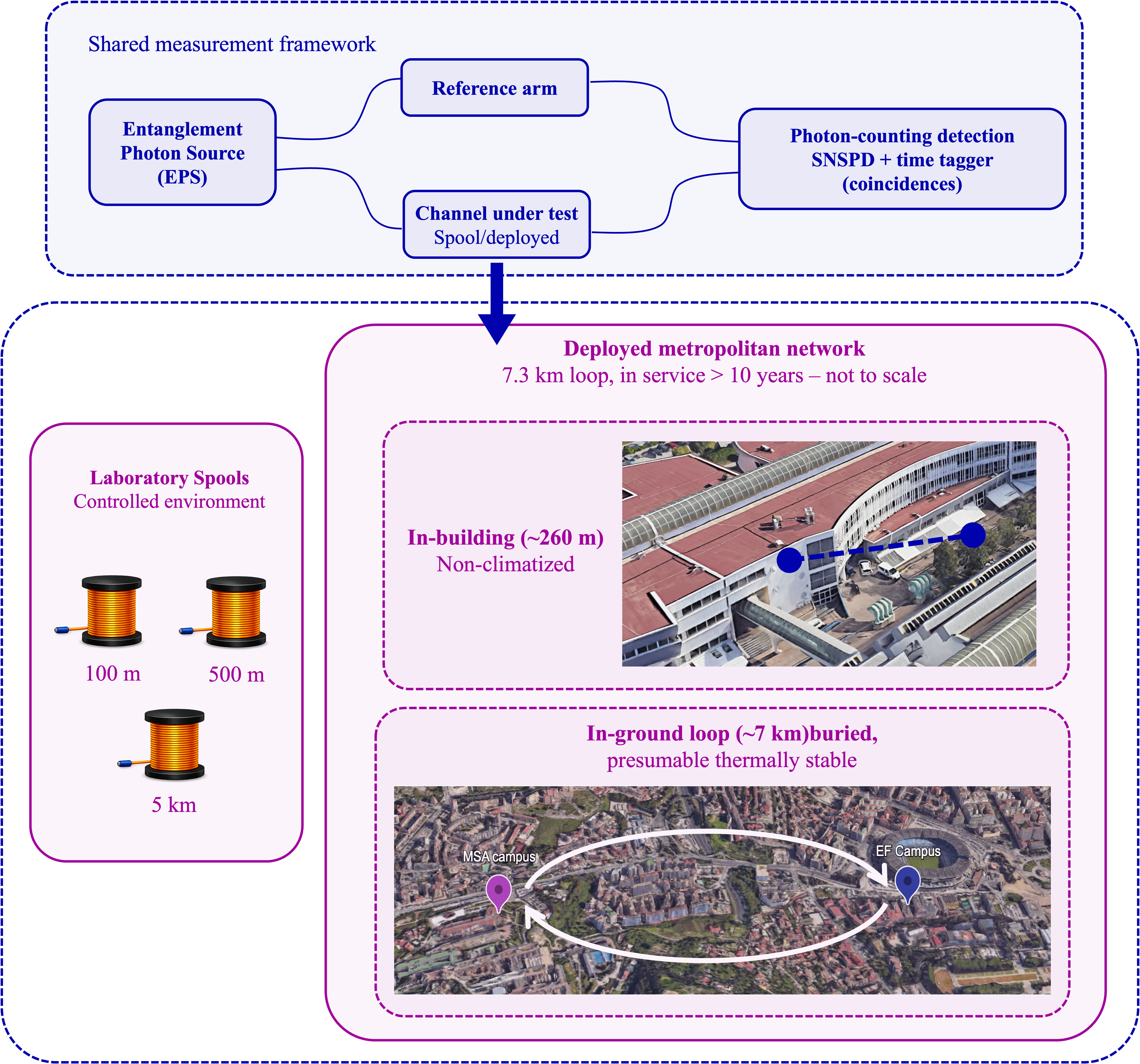} 
    \caption{Experimental framework (not to scale). (Top): the Entangled-Photon Source (EPS) emits polarization-entangled, temporally correlated photon pairs; one member (the so-called \textit{idler}) is sent to a local reference arm, while the other (the so-called \textit{signal}) probes the link under test. Their temporal coincidences are recorded by superconducting-nanowire single-photon (SNSPD) detectors read out by a time tagger, and feed the two channel metrics. Using one photon of each pair as a heralding reference enables coincidence-based suppression of background counts, providing the rationale for employing an entangled source rather than a bright classical probe. (Bottom): the channel under test is instantiated as either laboratory fiber spools ($100$~m and $5$~km, controlled environment) or the deployed metropolitan network: a $\sim 7.3$~km loop, in operational service for more than ten years, whose $\simeq 260$~m in-building segment runs through non-climatized corridors before reaching the buried in-ground loop connecting the Monte Sant'Angelo (MSA) and Engineering Faculty (EF) campuses of the University of Naples Federico II. Insets show geographic views of the in-building route and the in-ground loop. Measurement-specific configurations are detailed in Appendix~\ref{app:A}.}
    \hrulefill
    \label{fig:new:02}
\end{figure*}

All measurements reported in this work are obtained with the national \textit{QuantumInternet.it testbed} deployed at the University of Naples \textit{Federico II}, and summarized in Fig.~\ref{fig:new:02}. An Entangled-Photon Source (EPS) generates polarization-entangled photon pairs in the H/V basis, with internal DWDM filtering selecting the signal and idler wavelengths on the ITU grid. Of each pair, the idler is retained locally as a heralding reference, while the signal is launched into the channel under test; both are detected by SNSPDs and time-tagged. The detectors operate with a system detection efficiency of about $50\%$, a timing jitter lower than $100$~ps, and a dark-count rate lower than $100$~cps: figures of merit that set the noise floor against which the channel is characterized. Recording the two arms in coincidence, within a $1$ ns window well below the pair-generation period, suppresses the uncorrelated background. The residual measurement still includes the intrinsic quality of the entangled source and the detection electronics; a dedicated characterization of the source is deferred to a separate work~\cite{Preaparation}.

The channel under test is instantiated in two regimes. In the controlled regime, it is a laboratory fiber spool of either $100$~m or $5$~km, providing a stable baseline free of environmental perturbations. In the deployed regime, it is a $\simeq 7.3$km metropolitan loop that has been in operational service for more than ten years, connecting the Monte Sant'Angelo (MSA) and Engineering Faculty (EF) campuses of the University of Naples \textit{Federico II}. The loop is not a pristine link: it comprises a short in-building segment of about $\sim 260$~m spanning four floors through non-climatized corridors, followed by a buried in-ground section, and inherits the splices, connectors, and aging of infrastructure that already carries production traffic. It is precisely this real-world character, absent from laboratory spools, that the measurements in the following sections are designed to capture. The measurement-specific configurations used for each characterization are detailed in Appendix~\ref{app:A}.

\section{NOISE CHARACTERIZATION: QUANTUM SINR}
\label{sec:3}

Having described the setup, we now characterize the channel along the two metrics introduced in Sec.~\ref{sec:1.1}, beginning with the noise side.

\subsection{QSINR DEFINITION}
\label{sec:3.1}

A quantitative account of the noise side requires a metric that plays, for a quantum link, the role the SINR plays for a classical one. We adopt a photon-counting \textit{Quantum SINR} (QSINR), in which powers are replaced by photon counts. Photon-counting figures of merit of this kind have been introduced in the context of quantum-classical coexistence and routing\cite{BahElmCur-19, RuiGar-25}; here we adopt one such formulation and, crucially, populate its terms with measurements taken on deployed fiber.

To this aim, we define
\begin{equation}
    \label{eq:01}
    \mathrm{QSINR} = \frac{N_\mathrm{sig}}{N_\mathrm{n} + N_\mathrm{I}},
\end{equation}
where $N_\mathrm{sig}$ is the count of detected signal photons, $N_\mathrm{n}$ the intrinsic noise, and $N_\mathrm{I}$ the interference injected by co-propagating classical traffic. The signal term follows the same chain as a classical signal power:
\begin{equation}
    \label{eq:02}
    N_\mathrm{sig} = N_\mathrm{src}\,\eta_\mathrm{det}\,\eta_\mathrm{ch},
\end{equation}
with $N_\mathrm{src}$ denoting the photons emitted by the source\footnote{Which in turn depends primarily on the characteristics and the quality of the adopted quantum source, such as the generation probability and the fidelity of the generated quantum state. For a comprehensive overview of a rack-mountable entangled source characteristics and quality, we refer the reader to \cite{Preaparation}.}, $\eta_\mathrm{det}$ denoting the detection efficiency that depends primarily on the technology used in the detection system as shown in Table~\ref{tab:new:01}, and $\eta_\mathrm{ch}$ denoting the channel transmittance accounting for fiber attenuation, splices, and connectors.

The intrinsic noise $N_\mathrm{n}$ collects the contributions inherent to the receiver, primarily detector dark counts and additional background counts due to ambient photons (Sec.~\ref{sec:3.2}). The interference noise, instead, $N_\mathrm{I}$ collects the photons injected by classical traffic sharing the infrastructure, which can be classified into two contributions:
\begin{equation}
    \label{eq:03}
    N_\mathrm{I} = N_\mathrm{coex} + N_\mathrm{xt},
\end{equation}
where
\begin{itemize}
    \item $N_\mathrm{coex}$ denotes the interference generated by a classical signal co-propagating within the same optical fiber at a different frequency band (Sec.~\ref{sec:3.3}),
    \item $N_\mathrm{xt}$ denotes the inter-fiber crosstalk interference generated  by classical traffic propagating in adjacent fibers at the same frequency band of the quantum signal (Sec.~\ref{sec:3.4}).
\end{itemize}
Unlike a classical link -- where noise is dominated by receiver electronics -- our measurements show that the interference terms can exceed the intrinsic noise by orders of magnitude, and thus govern the achievable QSINR of a deployed quantum link.

\begin{table}[t]
\centering
\caption{Representative KPIs of the main single-photon detector technologies~\cite{DaoAmaAnd-24, WanYeKon-25}. The dark-count rate (DCR) sets the dominant contribution to the intrinsic noise $N_\mathrm{n}$ of Eq.~\eqref{eq:01}, while the detection efficiency $\eta_\mathrm{det}$ enters the detected-signal count of Eq.~\eqref{eq:02}.}
\label{tab:new:01}

\begin{tabular}{lcc}
\hline
\hline
\textbf{Detector} & \textbf{$\eta_\mathrm{det}$ (\%)} & \textbf{DCR (cps)}\\
\hline
Silicon SPAD & 55--80 & 10--300 \\
InGaAs SPAD & 20--35 & $10^{2}$--$2\times10^{3}$ \\
PMT & 15--35 & 20--500 \\
SNSPD & 90--99 & $<1$--10 \\
TES & 95--99 & $<10^{-3}$ \\
\hline
\hline
\end{tabular}

\begin{tablenotes}
\footnotesize
\item Acronyms: SPAD = Single-Photon Avalanche Diode; InGaAs = Indium Gallium Arsenide; PMT = Photomultiplier Tube; SNSPD = Superconducting Nanowire Single-Photon Detector; TES = Transition-Edge Sensor.
\end{tablenotes}

\end{table}

\subsection{INTRINSIC NOISE}
\label{sec:3.2}

The intrinsic noise $N_\mathrm{n}$ of Eq.~\eqref{eq:01} sets the floor below which no channel effect can be resolved. By definition, it collects the contributions present even when no classical traffic shares the infrastructure, and thus independent of the interference terms of Eq.~\eqref{eq:03}. We distinguish two such contributions: one originating in the receiver, the other in the surrounding environment.

The first is the detector dark-count rate (DCR), which for the SNSPDs employed here is below $100$~cps (Sec.~\ref{sec:2}), namely, two to three orders of magnitude lower than the SPAD-based receivers commonly used in fiber quantum communications (Table~\ref{tab:new:01}) and a direct consequence of the chosen detector technology.

The second contribution -- dominant when low-DCR detectors are employed -- is a background of ambient photons that couple into the fiber and reach the detector whenever the receiver remains connected to the deployed infrastructure, even with the quantum signal switched off. Measured on the metropolitan loop under these conditions, this background reaches $\sim 3 \times 10^{4}$~cps as shown in Fig.~\ref{fig:new:05} of Sec.~\ref{sec:3.3} -- more than two orders of magnitude above the detector dark counts -- and therefore constitutes the effective intrinsic-noise floor of a deployed link. Unlike the dark-count rate, which is fixed by the receiver, this term depends on how well the deployed fiber and its terminations are shielded from stray light, and can be mitigated through improved optical isolation at the receiving node.

\subsection{RAMAN INTERFERENCE}
\label{sec:3.3}

The first interference term, $N_\mathrm{coex}$ in Eq.~\eqref{eq:03}, arises from classical traffic co-propagating with the quantum signal in the same fiber. Since a fiber dedicated to quantum traffic alone as in \cite{CraLazPorSekFlaNam-24} is, at the network scale, economically unsustainable, integration with the classical infrastructure is unavoidable, and the two must share the same physical medium.

To keep the power disparity manageable, up to twelve orders of magnitude in our testbed, we allocate the classical traffic to the O-band ($1260-1360$~nm), a spectral region commonly used in conventional optical networks for short- and medium-distance communication links while reserving the lowest-attenuation C-band to the quantum signal as introduced in Sec.~\ref{sec:1.1}. Even with this spectral separation of almost $40$THz, spontaneous Raman scattering (SpRS) transfers a fraction of the classical photons across the spectrum, and some fall within the C-band, where they are indistinguishable from the quantum signal and constitute $N_\mathrm{coex}$.

Modeling this contribution is what turns coexistence from an obstacle into an engineering choice: a predictive model of the C-band Raman noise as a function of wavelength lets a designer select, within the ITU grid, the DWDM channel least affected by the co-propagating classical traffic. The Raman power generated at a C-band wavelength $\lambda_c$ by an O-band pump at $\lambda_o$ follows \cite{FroDynLuc-15,ThoKanKum-25,ThoKanXie-23}:
\begin{equation}
    \label{eq:04}
    P_\mathrm{Raman}(\lambda_o,\lambda_c) = \beta(\lambda_o,\lambda_c)\,P_o\,\Delta\lambda\,L_\mathrm{eff},
\end{equation}
where $P_o$ denotes the launched power, $\Delta\lambda$ the receiver filter bandwidth, $L_\mathrm{eff} = \left( e^{-\alpha_c L} - e^{-\alpha_o L} \right)/(\alpha_o - \alpha_c)$ the effective interaction length \cite{FroDynLuc-15,ThoKanKum-25} with $\alpha_o$ and $\alpha_c$ being the O- and C-band attenuation coefficients and $L$ being the fiber length, and $\beta(\lambda_o,\lambda_c)$ the Raman scattering coefficient, which carries the full wavelength dependence and is the quantity our model targets.

Since $P_o$, $\Delta\lambda$, and $L_\mathrm{eff}$ are known or measurable, the wavelength dependence is isolated in an effective coefficient extracted directly from the measured, power- and length-normalized click rate $C_\mathrm{SpRS,norm}$,
\begin{equation}
    \label{eq:05}
    \beta_\mathrm{eff}(\lambda_o,\lambda_c) = \frac{C_\mathrm{SpRS,norm}}{L_\mathrm{eff}},
\end{equation}
which removes the dependence on launched power and fiber length and yields the spectral shape of the Raman noise over the C-band.

\begin{figure}[t!]
    \centering
    \includegraphics[width=\linewidth]{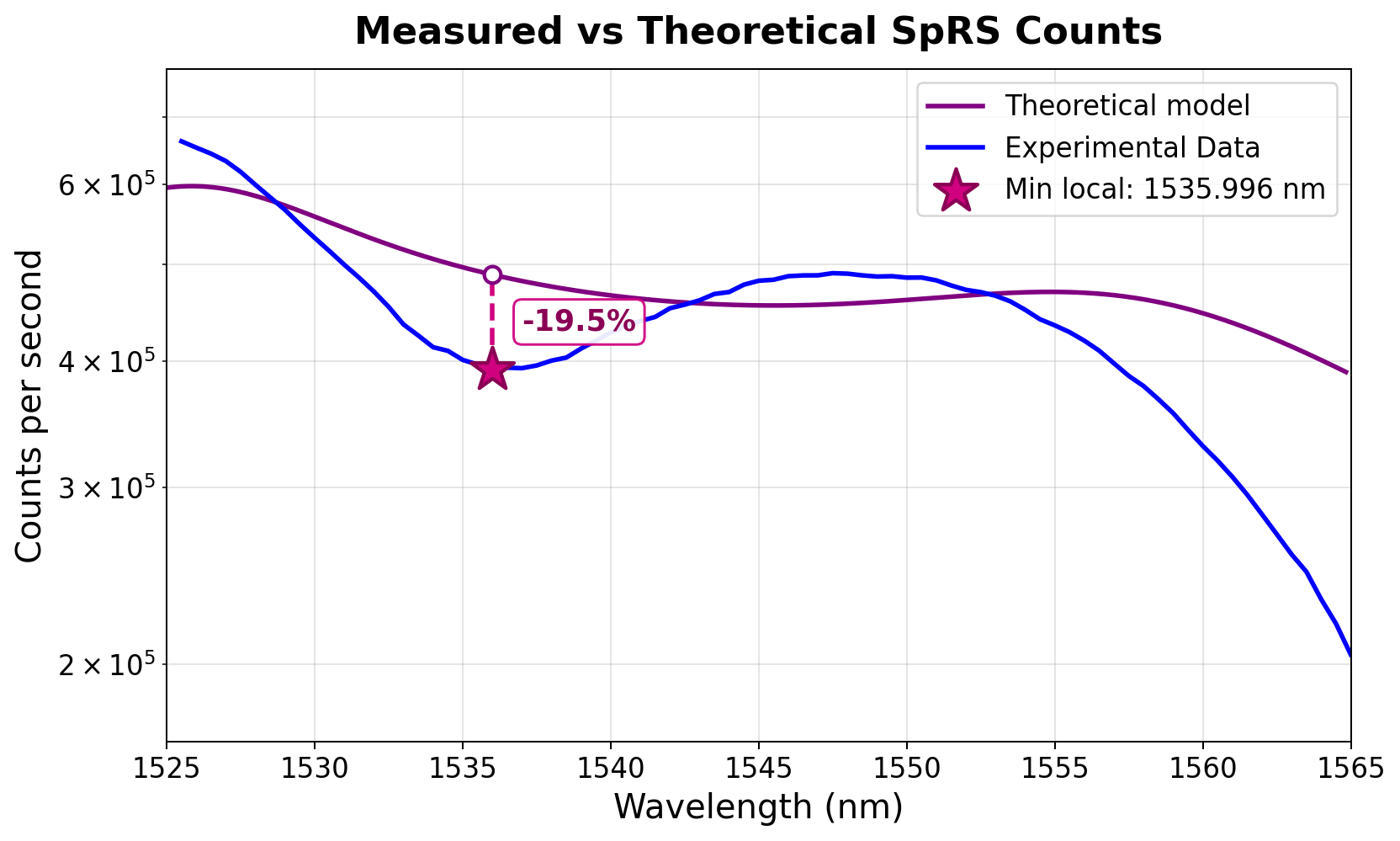}
    \caption{Measured C-band photon count rate generated by spontaneous Raman scattering from a classical signal co-propagating in the O-band at $1310$~nm, over the deployed SMF-28 fiber (blue), compared with the Hollenbeck-Cantrell prediction for undoped silica (purple) \cite{HolCan-02}. The measured spectrum exhibits a characteristic local minimum near $1535$~nm that the reference model does not reproduce, with the detected photon count departing from it by up to $19.5\%$. This minimum identifies the C-band operating point least affected by the co-propagating O-band traffic.}
    \label{fig:new:03}
    \hrulefill
\end{figure}

As a reference, the widely used Hollenbeck-Cantrell model \cite{HolCan-02} describes the Raman response of pure silica as a superposition of thirteen vibrational modes. Our central experimental finding is that the measured C-band spectrum departs from this reference in a specific, reproducible way: it exhibits a characteristic \textit{minimum} near $1535$~nm, absent from the original Hollenbeck-Cantrell prediction, from which the detected photon count departs by up to $19.5\%$ (Fig.~\ref{fig:new:03}).

This minimum is the \textit{operating point a network designer seeks}: allocating the quantum channel there minimizes the Raman noise injected by the co-propagating classical traffic at $1310$~nm, independently of any model used to describe the spectrum.

\begin{figure}[t]
    \centering
    \includegraphics[width=\columnwidth]{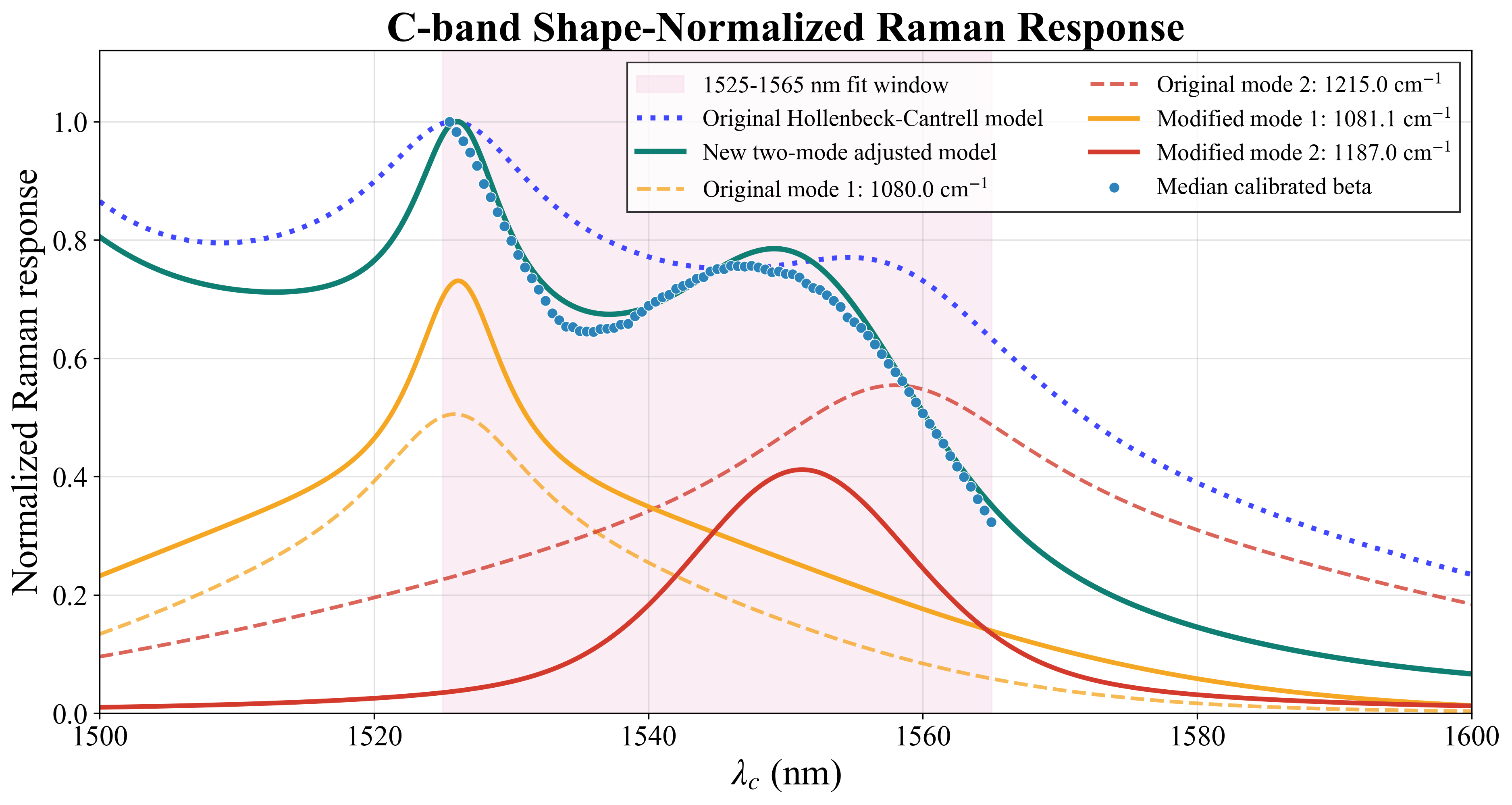}
    \caption{Normalized C-band Raman response: comparison between the experimental data measured across all laboratory sources and fiber lengths, the original Hollenbeck-Cantrell model for undoped silica~\cite{HolCan-02}, and the adjusted model obtained by modifying only the two highest-frequency Raman modes of the original model, namely modes 12 and 13. The individual contributions of modes 12 and 13 are also shown both in their original Hollenbeck-Cantrell form and as derived by the adjusted two-mode model, which follows the measured data closely, including the local minimum near $1535$~nm that is not reproduced by the original Hollenbeck-Cantrell response. All the curves are normalized to their respective maximum values within the 1525--1565 nm fitting window, highlighted in pink, in order to compare spectral shapes rather than absolute amplitudes.}
    \label{fig:new:04}
    \hrulefill
\end{figure}

To provide a compact description of this behavior, we proceed in two steps. Over the limited C-band interval accessible with our tunable filter ($\lambda_c \in [1525,1565]$~nm), a low-order (fourth-order) polynomial fit of the extracted $\beta_\mathrm{eff}$ already reproduces the measured shape \cite{DavMonGri-26} as detailed in Appendix~\ref{app:B}. Seeking a description with a physical rather than purely empirical basis, we then retain only the two highest-frequency modes of the Hollenbeck-Cantrell set -- modes 12 and 13, centered near $1080$ and $1215$~cm$^{-1}$, which are also the modes spectrally closest to the Raman shift relevant to our O-band-to-C-band configuration -- and re-fit their parameters to our data\footnote{Mode 12: center shifted from 1080.00 to 1081.09cm$^{-1}$, amplitude from 3.10 to 5.06, Gaussian width from 91.00 to 115.39cm$^{-1}$, and Lorentzian width from 30.33 to 17.15cm$^{-1}$. Mode 13: center shifted from 1215.00 to 1186.95cm$^{-1}$, amplitude from 3.40 to 2.85, Gaussian width from 160.00 to 32.92cm$^{-1}$, and Lorentzian width from 53.33 to 49.26cm$^{-1}$.} , obtaining the compact two-mode model shown together with the measured data and the reference curves in Fig.~\ref{fig:new:04}.

This two-mode model should be regarded as a phenomenological tool: it fairly reproduces the C-band Raman shape, but its parameters are fitted over a narrow spectral window and are not, on their own, uniquely identifiable, so we refrain from ascribing physical meaning to their individual values. A plausible reason for the departure from Hollenbeck-Cantrell is that the reference model is calibrated on undoped silica, whereas the deployed fiber is Ge-doped SMF-28; germanium doping is known to shift and reshape the vibrational bands of the silica network~\cite{HenBanFle-85, MicCorHen-06}, which could displace the local Raman response, and hence the minimum, from the undoped-silica case. Establishing this connection conclusively would require measuring $\beta$ across the full spectral range, beyond the tuning span of our filter, and is left to future work.

It is worthwhile to note that the predictive capability of this model was verified beyond the conditions under which it was derived. Using commercial transceivers rather than the narrowband-laser sources typical of controlled experiments, the model was tested against independent classical sources and fiber lengths not used in its identification, and against measurements performed directly on the deployed metropolitan loop. The adjusted model follows the experimental data across all laboratory sources and lengths (Fig.~\ref{fig:new:04}), and its predictions agree with the C-band Raman spectrum measured on the deployed inter-campus loop (Fig.~\ref{fig:new:05}); further validation against additional independent sources and lengths is reported in Appendix~\ref{app:B}. This consistency indicates that the model captures the underlying scattering mechanism rather than source- or length-specific features, and can therefore be used source-independently to guide Raman-aware channel allocation over deployed fiber.

\begin{figure*}[!t]
    \centering
    \subfloat[\textbf{Raw Data}\label{fig:RawData}]{%
        \includegraphics[width=0.48\textwidth]
        {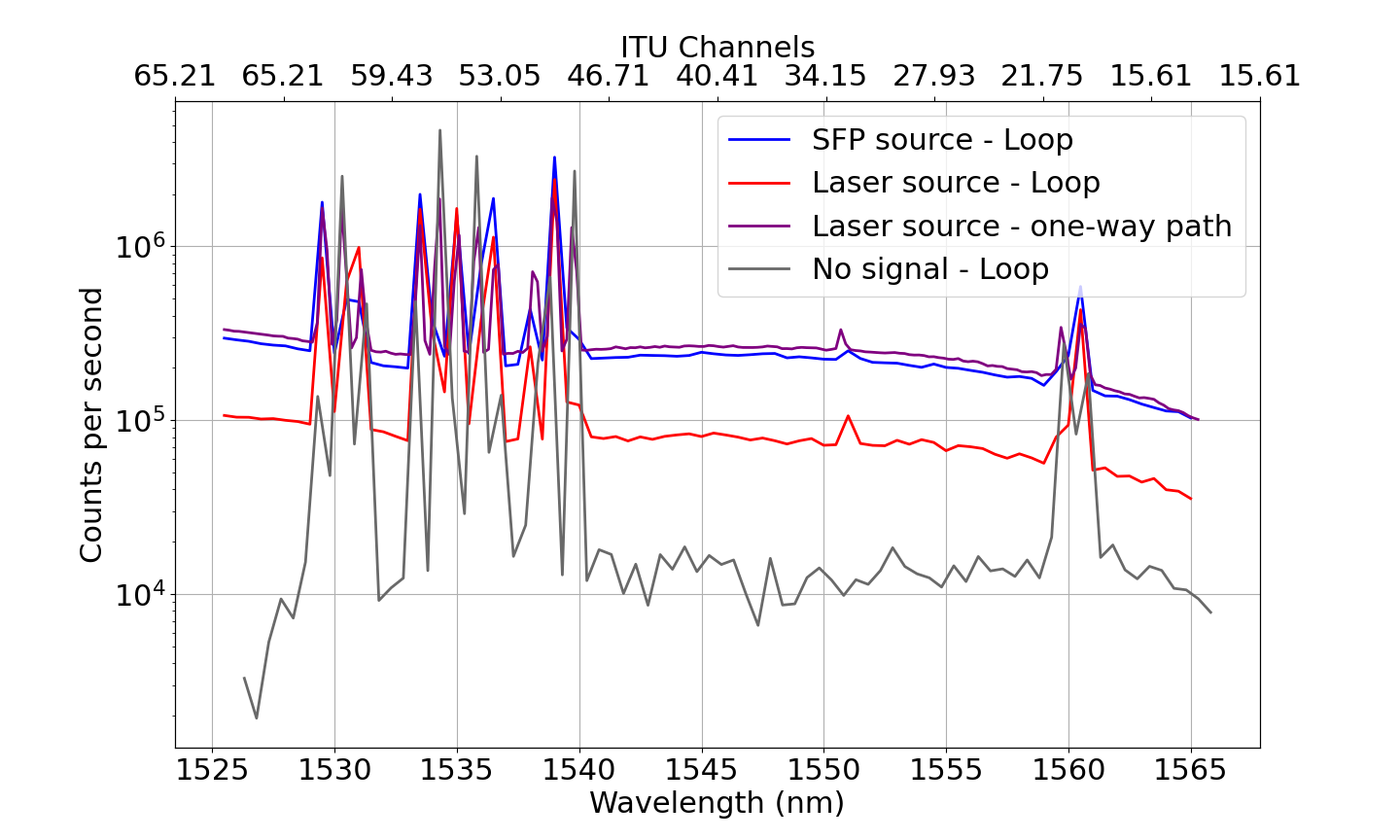}%
    }
    \hfill
    \subfloat[\textbf{Normalized Data}\label{fig:NormalizedData}]{%
        \includegraphics[width=0.48\textwidth]
        {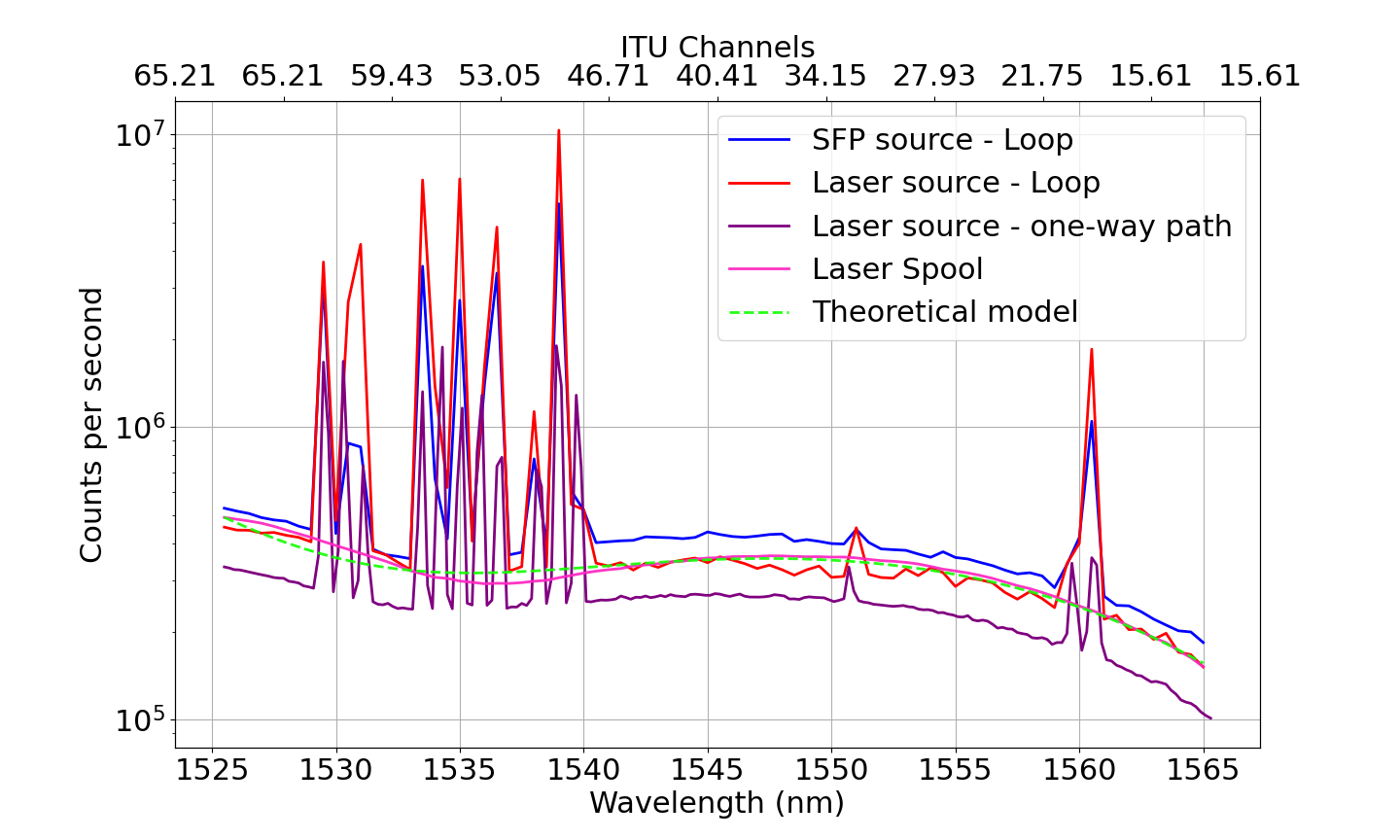}%
    }
    \caption{C-band SpRS photons generated by O-band classical sources
    as a function of wavelength, corresponding to the ITU DWDM channels.
    The measurements were obtained using both a narrowband laser source
    and a commercial Finisar SFP transmitter over both the complete loop
    (7~km) and a one-way path (3.5~km). (a) Raw data, including an additional reference measurement with no launched signal to identify background contributions. (b) Normalized data, including measurements obtained using an additional 5~km laboratory fiber spool. The results show consistency between the urban-scale and laboratory measurements after normalization.}
    \label{fig:new:05}
    \hrulefill
\end{figure*}

\subsection{CROSSTALK INTERFERENCE}
\label{sec:3.4}

The second interference contribution of Eq.~\eqref{eq:03}, $N_\mathrm{xt}$, arises from inter-fiber crosstalk. Unlike $N_\mathrm{coex}$, which is generated by classical traffic co-propagating with the quantum signal in the same fiber, $N_\mathrm{xt}$ originates from classical channels propagating in adjacent fibers of the same cable, allowing a fraction of their optical power to couple into the quantum fiber despite the physical separation.

Its signature is already evident in the deployed-loop measurements. In addition to the smooth Raman background described in Sec.~\ref{sec:3.3}, the spectra measured on the metropolitan loop (Fig.~\ref{fig:new:05}) exhibit several narrow spectral peaks, exceeding the surrounding photon counts by more than one order of magnitude, which are not explained by the Raman model.

Two observations identify the origin of these peaks. First, they occur at the wavelengths of the classical channels propagating in neighboring fibers rather than at Raman-shifted wavelengths; a representative example is the peak near $1560$~nm, corresponding to DWDM channels~21 and~22 used by the campus classical traffic. Second, as demonstrated by the dedicated verification experiment of Appendix~\ref{app:C}, the same spectral features are observed even when the monitored fiber carries no injected optical signal, ruling out any in-fiber generation mechanism.

To confirm that these counts are coupled from neighboring fibers, we performed a dedicated experiment in the in-building cable connecting the laboratory (floor $+2$) to the loop's point of presence (floor $-1$). A classical signal is launched into six fibers of the cable, patched in series to reach a propagation length of about $800$~m, while the spurious counts are measured on a separate pair of unfed fibers of the same cable (Appendix~\ref{app:C}). The counts detected on the idle fibers faithfully reproduce the spectral shape of the signal propagating in the active ones, confirming a coupling mechanism between co-located fibers. Because the deployed cable bundles several single-mode fibers within a shared jacket, classical traffic on any of them can leak into the fiber carrying the quantum signal, contributing to $N_\mathrm{xt}$.

Unlike the Raman term, which is a smooth, broadband background amenable to the wavelength-allocation strategy of Sec.~\ref{sec:3.3}, crosstalk is narrowband and localized at the classical carriers' wavelengths, and is therefore mitigated instead by spectral filtering and by an informed choice of which fibers in the cable carry quantum and classical traffic. The numerical values of these noise contributions, together with the Raman and intrinsic-noise terms, are collected in the link-budget summary of Sec.~\ref{sec:5}.

\section{QUANTUM-STATE DEGRADATION CHARACTERIZATION}
\label{sec:4}

Having characterized the noise present and the interference injected into the quantum link, we now turn to the second metric: the degradation of the quantum state that the information carrier -- namely, the photon -- transports. Whereas the noise characterization assesses \textit{how many} spurious photons reach the detector, the degradation characterization assesses \textit{how faithfully} the state carried by a quantum-signaling photon survives propagation.

Quantum information can be encoded in any of several photonic degrees of freedom; the three most widely used -- polarization, time, and frequency -- respond differently to the perturbations of a deployed fiber. Rather than a single figure of merit, the degradation side therefore requires a per-degree characterization: for each encoding we quantify both the channel-induced distortion of the state and, crucially, its \emph{drift over time}, since a metropolitan link is not stationary but evolves under thermal and mechanical stress.

After describing the measurement approach common to all three degrees of freedom (Sec.~\ref{sec:4.1}), we treat polarization (Sec.~\ref{sec:4.2}), time (Sec.~\ref{sec:4.3}), and frequency (Sec.~\ref{sec:4.4}) in turn, and collect the resulting parameters into the link budget of Sec.~\ref{sec:5}.

\subsection{CHARACTERIZATION RIGOR AND ACCURACY}
\label{sec:4.1}
Throughout, the entangled pair serves as a probe. One photon, the \textit{idler} in Fig.~\ref{fig:new:02}, is retained locally over a short ($\sim 10$~m) reference path and plays two roles. First, it provides a nearly stationary temporal and state reference against which the \textit{signal} photon, after traversing the channel under test, is compared, so that the recovered correlations isolate how the channel has acted on the transmitted state. Second, recording the two photons in coincidence suppresses the dark counts and, crucially, the classical-traffic interference characterized in Sec.~\ref{sec:3}. It does not remove the source and detector contributions -- the finite fidelity of the generated state and the detection jitter -- which set a measurement floor independent of the channel; the source characterization is reported separately~\cite{Preaparation}.

To ensure that differences among the channels reflect the channels themselves rather than the instant at which each is probed, the signal photon is split by a $1{:}4$ beam splitter and launched \emph{simultaneously} into the three quantum links under study -- the $100$~m and $5$~km laboratory spools and the $7.3$~km deployed inter-campus loop -- which are therefore characterized concurrently under identical source and environmental conditions. Periodic source realignment ensures that the source remains polarization-stable throughout the measurement campaign, so that the observed polarization evolution can be attributed to the transmission channel rather than to the entangled-photon source. The measurement configuration common to all three degrees of freedom is detailed in Appendix~\ref{app:A}. 

\begin{remark}
    The coincidence-based characterization should be interpreted as the reference measurement of the intrinsic evolution of the quantum signal, since the coincidence gate effectively rejects detector noise and photons generated by co-propagating classical traffic. In contrast, practical prepare-and-measure systems, such as BB84, do not have access to a heralding photon and therefore estimate the channel polarization from the detected signal events alone. As a result, the estimated polarization drift is influenced not only by the quantum signal but also by detector noise and classical-traffic interference, corresponding to the drift observed in the singles measurements.
\end{remark}

The characterizations reported in this section follow a consistent methodology: for each degree of freedom, we report a representative acquisition and verify that its qualitative behavior is reproduced across multiple independent runs. While the absolute values naturally vary between acquisitions, as expected for a channel that evolves under changing environmental conditions, the phenomenology is stable: the same channels consistently emerge as the most and least perturbed along each degree of freedom. The specific runs are identified in the respective subsections.

\subsection{POLARIZATION DEGRADATION}
\label{sec:4.2}

\begin{figure*}[t]
    \centering
    \begin{minipage}{1\textwidth}
        \centering
        \includegraphics[width=\textwidth]{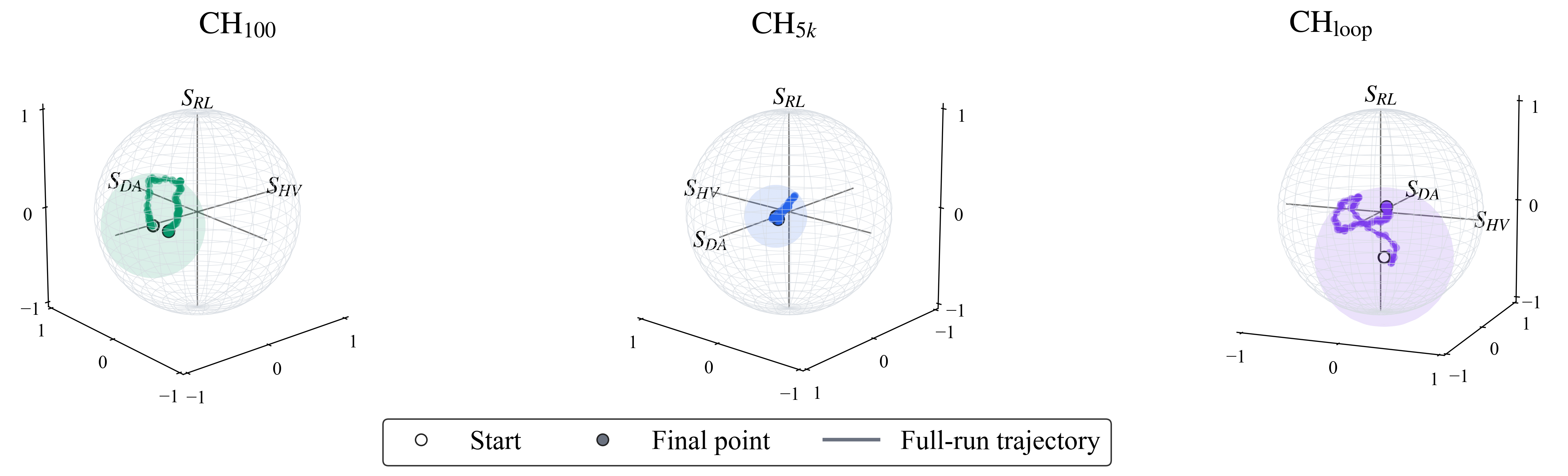}
        \caption{Temporal evolution of the polarization state reconstructed from singles measurements during the $\sim 14$-hour acquisition. Since the reconstruction is based solely on signal-detector counts, it includes the contribution of detector noise and classical-traffic interference. The reconstructed trajectory is most compact on the laboratory spools and broadest on the deployed loop.}
        \label{fig:new:06}
        \hrulefill
    \end{minipage}
    \vspace{6pt}
    \begin{minipage}{1.0\textwidth}
        \centering
        \includegraphics[width=\textwidth]{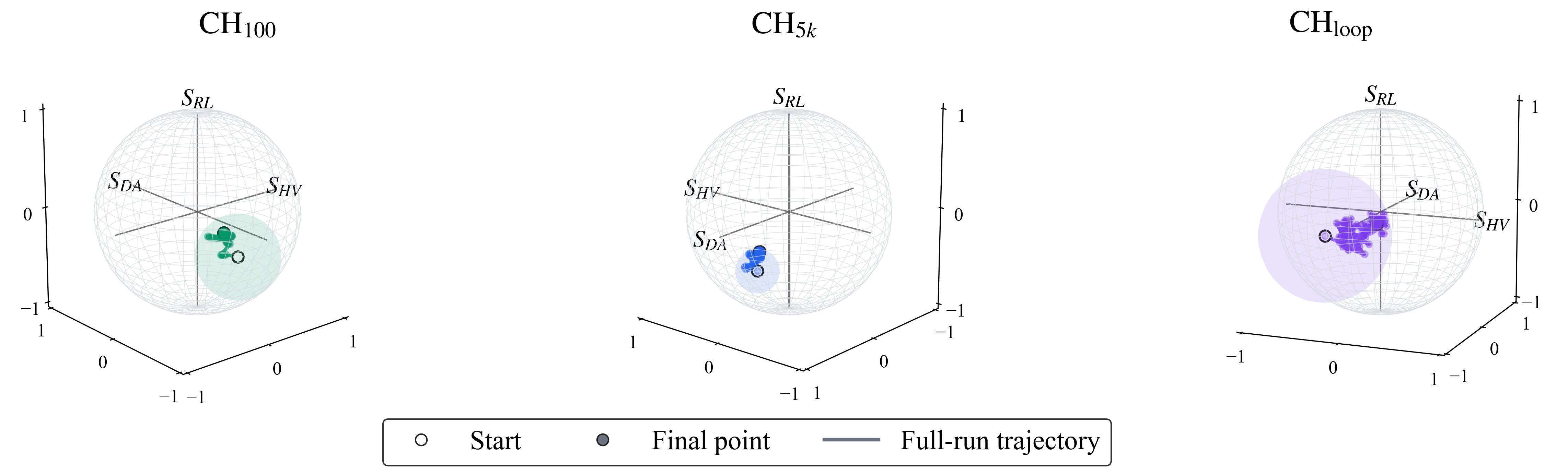}
        \caption{Evolution of the polarization state reconstructed from coincidence measurements during the ~14-hour acquisition. Compared with the singles reconstruction of Fig.~\ref{fig:new:06}, the trajectories are markedly more compact on the laboratory spools, while the deployed loop still spans a broad region.}
        \label{fig:new:07}
        \hrulefill
    \end{minipage}
\end{figure*}

Polarization is the most common encoding\footnote{Since it can be prepared, manipulated with high precision, and measured using relatively simple optical components, such as polarizers or polarization controllers.}  for photonic qubits, but also the most exposed to the fiber environment: birefringence induced by mechanical stress, bending, and temperature continuously rotates the polarization state as it propagates, so that a state prepared as $\ket{H}$ may arrive as an arbitrary one. On a deployed link this rotation is not fixed but drifts in time, and the relevant question for a network designer is not only how much the polarization is disturbed, but how fast that disturbance evolves relative to the timescale of a quantum protocol.

Assessing the polarization drift is indeed crucial, since it directly affects the \textit{fidelity} of the distributed state. And fidelity is the key performance indicator for any quantum network \cite{RFC9340}, as it governs the quality of every downstream protocol, from teleportation to entanglement swapping and quantum key distribution.

To quantify the degradation, we reconstruct the polarization state emerging from the channel by quantum state tomography\footnote{Quantum state tomography (QST) reconstructs the density matrix of an unknown quantum state from measurements performed on multiple identically prepared copies of the state~\cite{AltJefKwi-05}. For polarization qubits, the reconstructed density matrix directly determines the corresponding point on the Poincar\'e sphere. Here a \emph{single} input state is prepared (Appendix~\ref{app:A.4}), so the results characterize the channel's action on that state rather than the full process: unlike process tomography, they are specific to the chosen input polarization.} (the acquisition procedure is detailed in Appendix~\ref{app:A.4}), and we track its evolution on the Poincar\'e sphere\footnote{The Poincar\'e sphere is mathematically equivalent to the Bloch sphere commonly used for representing qubit states~\cite{CacCalVan-20}, differing only by the choice of coordinate axes.} over the $\sim 14$-hour acquisition\footnote{Acquisition run \texttt{260619\_163039}, in the format \texttt{YYMMDD\_HHMMSS}.}.

\begin{table}[t]
    \centering
    \caption{Maximum polarization displacement, $r_{\max}$, evaluated with respect to the initial reconstructed state for single-photon and coincidence-based polarization tomography.}
    \label{tab:new:04}
    \begin{tabular}{lcc}
        \hline
        \hline
        \textbf{Channel} & \textbf{Single-photon tomography} & \textbf{Coincidence-based tomography} \\
        \hline
        \hline
        $\mathrm{CH}_{100}$ & 0.515 & 0.438 \\
        $\mathrm{CH}_{5k}$ & 0.304 & 0.209 \\
        $\mathrm{CH}_{\mathrm{loop}}$ & 0.676 & 0.637 \\
        \hline
        \hline
    \end{tabular}
\end{table}

As a first step, Fig.~\ref{fig:new:06} reports the temporal evolution of the polarization states reconstructed from the singles measurements, by periodically repeating the tomographic reconstruction throughout the experiment. Since these are not conditioned on the heralding idler photon, they include the detector noise and classical-traffic background, and thus reflect the state seen by a practical prepare-and-measure receiver (cf.\ Remark~1) rather than the intrinsic channel evolution.
\begin{figure}[t]
    \centering
    \includegraphics[width=0.9\columnwidth]{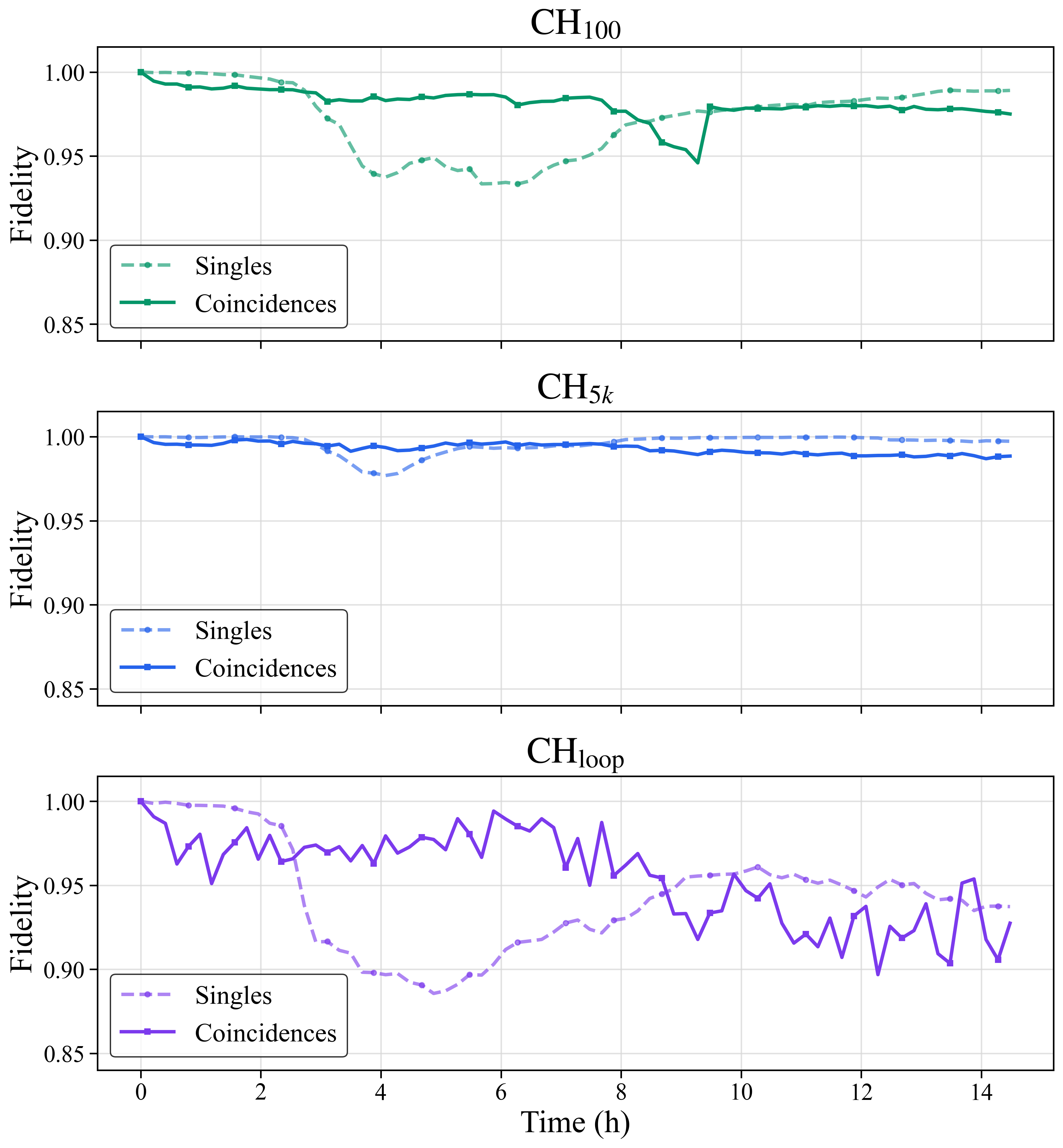} 
    \caption{Polarization fidelity with respect to the initial reconstructed state as a function of time for the three analyzed channels. Dashed and solid curves correspond to singles and coincidence tomography, respectively.}
    \label{fig:new:08}
    \hrulefill
\end{figure}

The reconstructed trajectories reveal markedly different behaviors across the three channels. The laboratory spools ($\mathrm{CH}_{100}$ and $\mathrm{CH}_{5k}$) exhibit relatively smooth polarization evolution, whereas the deployed loop spans a substantially larger region of the Poincar\'e sphere, reflecting the stronger environmental perturbations experienced by the metropolitan fiber. Since singles tomography is directly affected by detector dark counts and by photons generated by classical coexistence, the reconstructed trajectories combine the intrinsic channel dynamics with measurement-induced fluctuations. 

To isolate the intrinsic polarization evolution from the measurement background, the same analysis is repeated using coincidence-based tomography, which significantly suppresses uncorrelated detection events, yielding a more faithful reconstruction of the transmitted quantum state. This effect is particularly evident for the laboratory channels, whose trajectories become substantially more compact while preserving the underlying polarization evolution. By contrast, the deployed loop continues to span a broad region of the Poincar\'e sphere, indicating that the dominant contribution arises from genuine channel-induced polarization drift rather than measurement noise. Because coincidence tomography relies on a much smaller subset of the recorded events, longer integration times are required to achieve reliable state reconstruction. By contrast, the deployed loop continues to span a broad region of the Poincaré sphere, indicating that the dominant contribution arises from genuine channel-induced polarization drift rather than measurement noise. Because coincidence tomography relies on a much smaller subset of the recorded events, longer integration times are required to achieve reliable state reconstruction. Accordingly, Fig.~\ref{fig:new:07} shows temporal evolution of the polarization states reconstructed from the coincidences measurements.

To summarize these observations with a single scalar metric, we define $r_{\max}$ as the maximum Euclidean distance between the reconstructed polarization state and its initial position on the Poincar\'e sphere during the acquisition. The corresponding values are reported in Table~\ref{tab:new:04}.

Coincidence filtering reduces $r_{\max}$ for both laboratory channels, with the largest improvement observed for $\mathrm{CH}_{5k}$, confirming that a significant fraction of the apparent polarization fluctuations originates from uncorrelated background events. For the deployed loop, the reduction is considerably smaller, indicating that the dominant contribution is the intrinsic time-varying birefringence of the metropolitan fiber. Consequently, coincidence filtering improves the accuracy of the reconstruction but cannot compensate for the underlying polarization dynamics, which ultimately require active stabilization.

\begin{figure}[t]
    \centering
    \includegraphics[width=0.9\columnwidth]{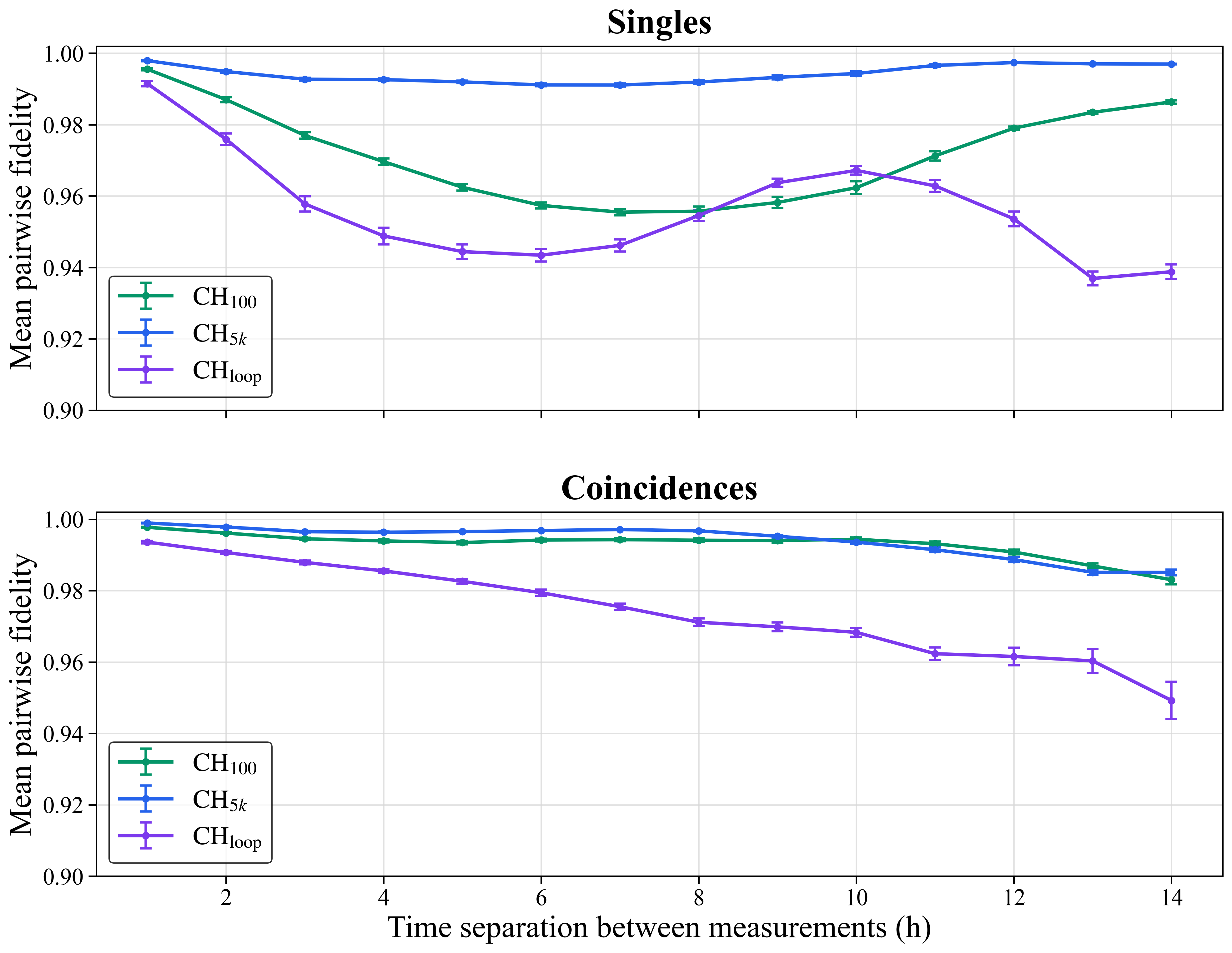} 
    \caption{Mean pairwise polarization fidelity as a function of the temporal separation between measurements. Each point is obtained by averaging the fidelity over all pairs of reconstructed states separated by the same time interval. Error bars denote the standard error of the mean.}
    \label{fig:new:09}
    \hrulefill
\end{figure}

The polarization drift over time can be further quantified through the state fidelity. Fig.~\ref{fig:new:08} reports the fidelity of each reconstructed state with respect to the initial polarization state, providing a measure of the cumulative drift from the initial calibration. Fig.~\ref{fig:new:09} instead reports the mean pairwise fidelity between all reconstructions separated by the same temporal interval. Unlike the fidelity to the initial state, the pairwise metric directly characterizes the rate at which the polarization state evolves, independently of the initial reference, and therefore provides the relevant engineering quantity for determining how frequently polarization compensation should be updated. The corresponding summary statistics are reported in Table~\ref{tab:new:05}.

The laboratory channels maintain high polarization fidelity throughout the acquisition, and the singles- and coincidence-based values nearly coincide. Indeed, on the spools the background is limited to dark counts, with no interference to reject, so coincidence gating has little effect and the small residual differences reflect the larger statistical uncertainty of the smaller coincidence sample. For $\mathrm{CH}_{5k}$, both reconstruction methods yield fidelities consistently above $0.99$, confirming the excellent polarization stability of the laboratory fiber. The slightly larger displacement observed on $\mathrm{CH}_{100}$ compared with $\mathrm{CH}_{5k}$ does not reflect a stronger channel perturbation, since both are stable laboratory spools. We attribute it to residual differences in the tomographic accuracy of the independent polarization analyzers used on the two beam-splitter arms.

The deployed metropolitan loop exhibits a markedly different behavior. Coincidence filtering improves the minimum and mean fidelity (from  $0.942$ to $0.956$) by removing the Raman and crosstalk background. The lower final value, however, reflects the increased statistical uncertainty of the coincidence reconstruction, whose count rate is strongly reduced by the loop attenuation and by the narrowband filtering applied on that link. Beyond these statistical effects, both the temporal evolution in Fig.~\ref{fig:new:08} and the pairwise analysis in Fig.~\ref{fig:new:09} reveal a progressive de-correlation of the polarization state over time, showing that the dominant impairment is the channel's own time-varying birefringence rather than measurement noise.

\begin{table}[t]
    \centering
    \caption{Minimum, mean, and final polarization fidelity evaluated against the initial reconstructed state for singles and coincidence tomography.}
    \label{tab:new:05}
    \begin{tabular}{lccc}
        \hline
        \hline
        Channel and mode & $F_{\min}$ & $\overline{F}$ & $F_{\mathrm{final}}$ \\
        \hline
        \hline
        $\mathrm{CH}_{100}$ singles & 0.933 & 0.973 & 0.989 \\
        $\mathrm{CH}_{100}$ coincidences & 0.946 & 0.981 & 0.975 \\
        $\mathrm{CH}_{5k}$ singles & 0.977 & 0.996 & 0.997 \\
        $\mathrm{CH}_{5k}$ coincidences & 0.987 & 0.993 & 0.988 \\
        $\mathrm{CH}_{\mathrm{loop}}$ singles & 0.886 & 0.942 & 0.937 \\
        $\mathrm{CH}_{\mathrm{loop}}$ coincidences & 0.897 & 0.956 & 0.927 \\
        \hline
        \hline
    \end{tabular}
\end{table}

While the fidelity with respect to the initial state remains high for all channels throughout the measurement campaign, the pairwise fidelity reveals markedly different temporal dynamics. Both laboratory spools preserve a high pairwise fidelity even over the longest observation intervals, indicating that polarization evolves only slowly in controlled environments. By contrast, the deployed metropolitan loop exhibits a progressive decrease of the pairwise fidelity with increasing temporal separation, demonstrating that the polarization state continuously decorrelates under realistic environmental perturbations.

This distinction has a direct engineering implication. The fidelity to the initial state quantifies the cumulative degradation of a calibrated link, whereas the pairwise fidelity determines the characteristic time scale over which the polarization state loses correlation. The latter therefore provides the appropriate metric for selecting the update interval of polarization-compensation algorithms, in the same spirit that temporal-drift measurements determine the recalibration interval for synchronization. This consideration is incorporated into the link-budget analysis of Sec.~\ref{sec:5}.

\subsection{TEMPORAL DEGRADATION}
\label{sec:4.3}

The second degree of freedom commonly used to encode quantum information is timing, exploited for instance through time-bin encoding of DV qubit states. Encoding in this degree of freedom offers several advantages, since the information is associated with the photon's arrival in a specific time slot, and its detection is essentially based on the discrimination between the photon's presence and absence within predefined time intervals. Besides time-bin encoding, precise temporal synchronization is a fundamental requirement in several quantum communication protocols. A notable example is all-photonic entanglement swapping \cite{PanBouWei-98}, where high-visibility Hong-Ou-Mandel interference \cite{HonOuMan-87} requires photons arriving at the Bell-state measurement to be temporally indistinguishable. Consequently, even small channel-induced delay variations may degrade the interference visibility and compromise the protocol~\cite{DavCacCal-26}.

Over a deployed link, however, the optical path length is not constant: thermal expansion and mechanical stress continuously change the fiber's effective length, and hence the propagation delay of the signal relative to the reference. On a metropolitan link this delay does not merely offset the coincidence peak once, but drifts throughout operation, progressively displacing the signal photons from the window in which they are expected.

We quantify this effect by measuring, over the same 
14-hour acquisition\footnote{Acquisition run \texttt{260619\_163039}.} used for the polarization characterization, the arrival-time difference\footnote{The two entangled photons are generated essentially simultaneously at the source, but reach their detectors after a propagation delay determined by the channel length; in standard single-mode fiber this delay is approximately $35$--$40$~ps/km.} $\tau_i$ between each signal photon propagating through channel $i$ and the locally retained idler photon. As detailed in Appendix~\ref{app:A.4}, at each acquisition step $n$ the temporal cross-correlation histogram between the signal and idler detections is fitted with a Gaussian function, whose mean provides the estimate $\tau_i(n)$ of the propagation delay. We then track both its \textit{drift} -- defined as the variation of the fitted Gaussian mean relative to its initial value,
\begin{equation}
    \Delta\tau_i(n) = \tau_i(n) - \tau_i(0),
    \label{eq:drift}
\end{equation}
where $\tau_i(0)$ denotes the delay at the beginning of the acquisition -- and its \textit{dispersion}, quantified by the Gaussian standard deviation of the same correlation peak. Together, these quantities characterize the long-term temporal stability of the quantum link.

\begin{figure}[t]
    \centering
    \includegraphics[width=0.9\columnwidth]{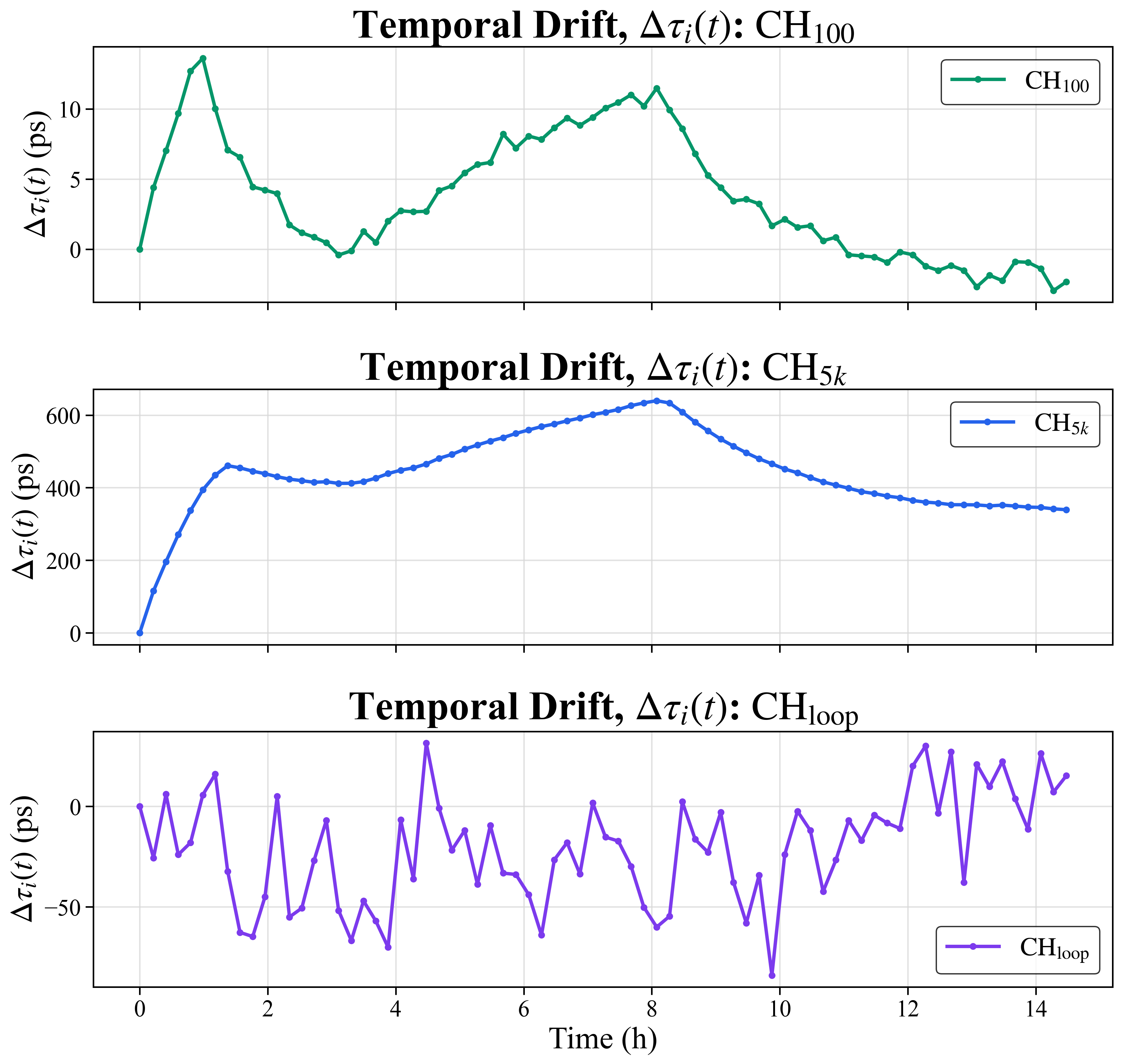} 
    \caption{Temporal drift of the relative delay between the reference channel ($\mathrm{CH}_{\mathrm{ref}}$) and each channel under test over the 14-hour acquisition. At each acquisition step, the delay is estimated as the mean of the Gaussian fitted to the temporal cross-correlation histogram (Appendix~\ref{app:A.4}. The reported drift corresponds to the variation of this fitted mean with respect to its initial value.}
    \label{fig:new:10}
    \hrulefill
\end{figure}

Figure~\ref{fig:new:10} reports the temporal evolution of $\Delta\tau_i(t)$ for the three channels. The two laboratory spools exhibit a remarkably similar behavior: despite their different lengths, the propagation delay evolves smoothly over time, following comparable monotonic increases and decreases throughout the acquisition. This suggests that both fibers experience the same slowly varying laboratory environment. In contrast, the deployed inter-campus loop no longer exhibits such regularity. Instead, the propagation delay fluctuates irregularly around its initial value, consistent with the larger and less predictable environmental perturbations experienced by the outdoor infrastructure.

\begin{figure}[t]
    \centering
    \includegraphics[width=0.9\columnwidth]{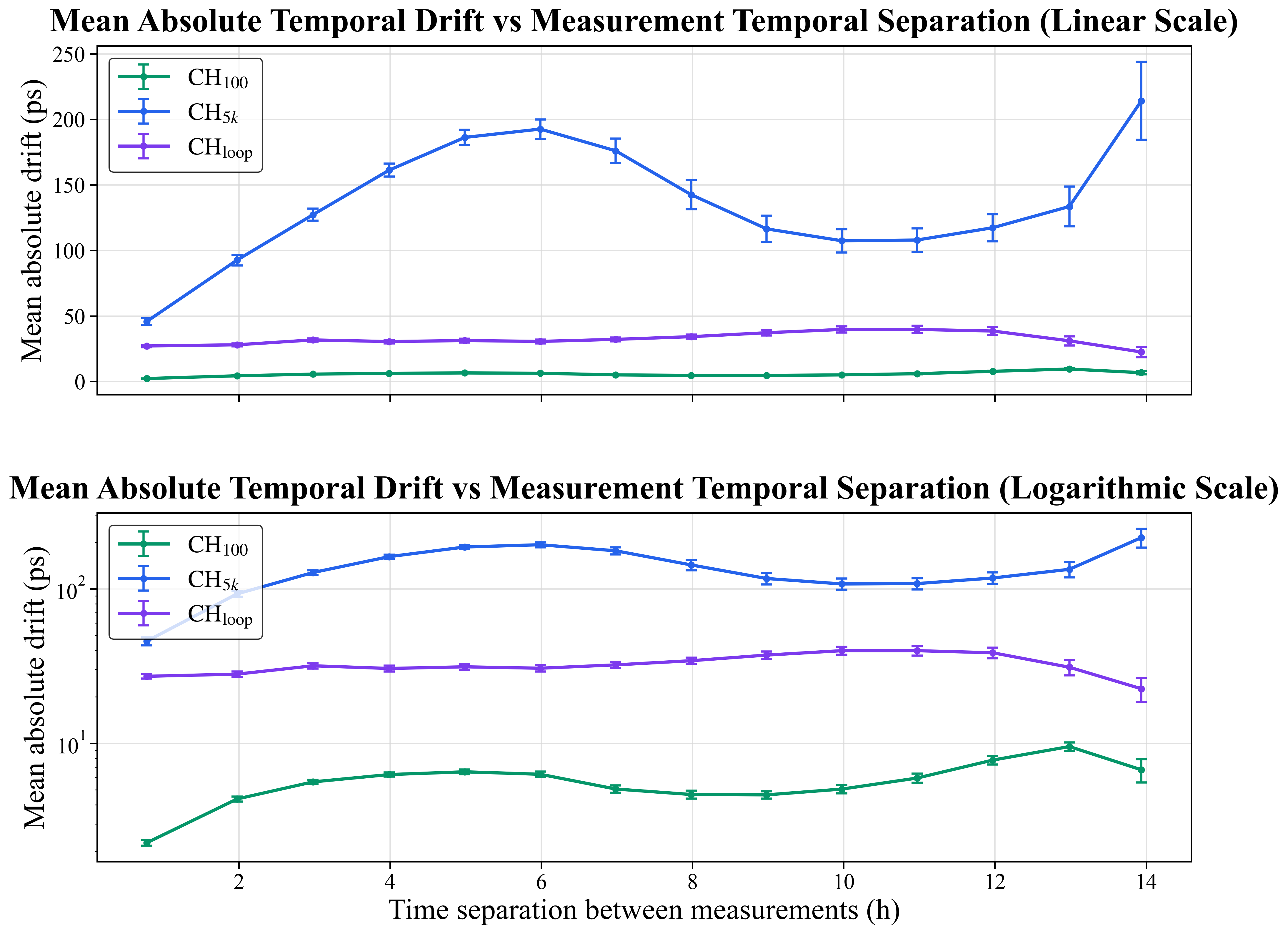} 
    \caption{Mean absolute temporal drift as a function of the time separation between measurements. For each channel and each measurement separation $\Delta t$, the absolute drift $|\Delta\tau_i(t+\Delta t)-\Delta\tau_i(t)|$ is computed over all available pairs of measurements and subsequently averaged. Error bars denote the standard error of the mean. The same data are shown on linear and logarithmic scales to emphasize both the absolute drift magnitude and the relative differences among the three channels.}
    \label{fig:new:11}
    \hrulefill
\end{figure}

To provide a statistical characterization of the temporal stability, Fig.~\ref{fig:new:11} reports the mean absolute temporal drift as a function of the time separation between measurements. For each separation $\Delta t$, all pairs of measurements separated by $\Delta t$ are considered, the corresponding absolute drifts are computed, and their mean is reported; the error bars indicate the standard error of the mean. The $100$~m laboratory spool exhibits the smallest delay variations, remaining below $10$~ps throughout the entire observation interval. The $5$~km laboratory spool experiences substantially larger fluctuations, reaching approximately $200$~ps after several hours. Despite this difference in magnitude, both laboratory spools exhibit a similar dependence on the measurement separation, with the mean drift increasing over the first hours, reaching a maximum after approximately $5$–$6$~hours, and subsequently decreasing. By contrast, the deployed inter-campus loop shows a markedly different behavior, with comparatively stable delay variations around $30$–$40$~ps over the entire acquisition. These results indicate that temporal stability depends not only on the fiber length but also on the environmental conditions and installation characteristics of the link. Notably, the buried metropolitan fiber exhibits smaller pairwise drift than the longer laboratory spool, suggesting that the thermally stable underground environment provides better temporal isolation than the laboratory. This finding has been confirmed across multiple independent measurement campaigns not reported here for brevity.

\begin{figure}[t]
    \centering
    \includegraphics[width=0.9\columnwidth]{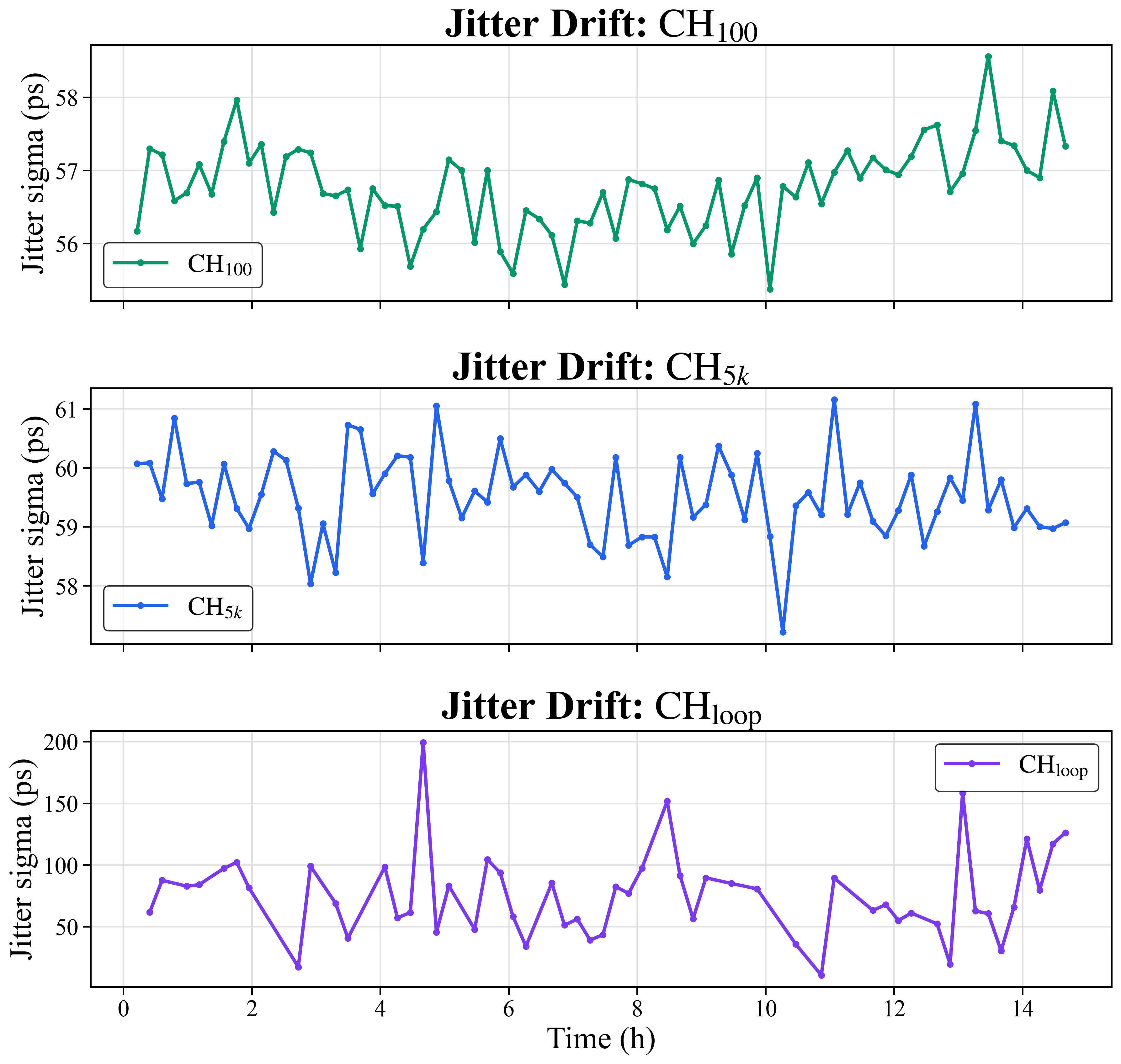} 
    \caption{Temporal correlation jitter over the 14-hour acquisition, estimated as the Gaussian standard deviation of the fitted signal-idler temporal cross-correlation peak.}
    \hrulefill
    \label{fig:new:12}
\end{figure}

While Figs.~\ref{fig:new:10}-\ref{fig:new:11} characterize the evolution of the propagation delay, the temporal quality of the correlation peak is analyzed next through its Gaussian width. More into details, the Gaussian standard deviation extracted from the temporal cross-correlation is reported in Fig.~\ref{fig:new:12} as a measure of the correlation-peak width. For both laboratory spools, this quantity remains nearly constant throughout the experiment, with values around $55$–$60$~ps. Such values are consistent with the intrinsic timing resolution of the measurement system, dominated by the quadratic combination of the timing jitter of the two SNSPDs (specified below $100$~ps FWHM) together with the $8$~ps contribution of the time tagger. In contrast, the deployed inter-campus loop exhibits a substantially broader and more unstable correlation peak, with occasional Gaussian standard deviations approaching $200$~ps. This additional broadening indicates that the deployed channel introduces temporal fluctuations beyond the instrumental contribution, thereby reducing the temporal indistinguishability of the detected photon pairs.

Together, these two quantities provide complementary information on the temporal behavior of the quantum link: the former determines the synchronization requirements, whereas the latter quantifies the preservation of temporal indistinguishability. Yet, of the two, the temporal drift has the most direct operational consequence. In practical quantum communication systems synchronization cannot be performed once and left unchanged; instead, it must be periodically repeated whenever the accumulated delay exceeds the timing tolerance of the protocol. Resynchronization is therefore required whenever
\begin{equation}
    |\Delta\tau_i(t)| \geq \tau_\mathrm{th},
    \label{eq:recal}
\end{equation}
where $\tau_{\rm th}$ is the maximum admissible timing error. In general, this threshold is determined by the coincidence window or timing tolerance of the protocol under consideration. The temporal-drift statistics reported in Fig.~\ref{fig:new:11} therefore provide a direct estimate of the maximum interval between successive synchronization procedures for a given application.

\subsection{FREQUENCY DEGRADATION}
\label{sec:4.4}

\begin{figure} 
    \centering
    \includegraphics[width=\linewidth]{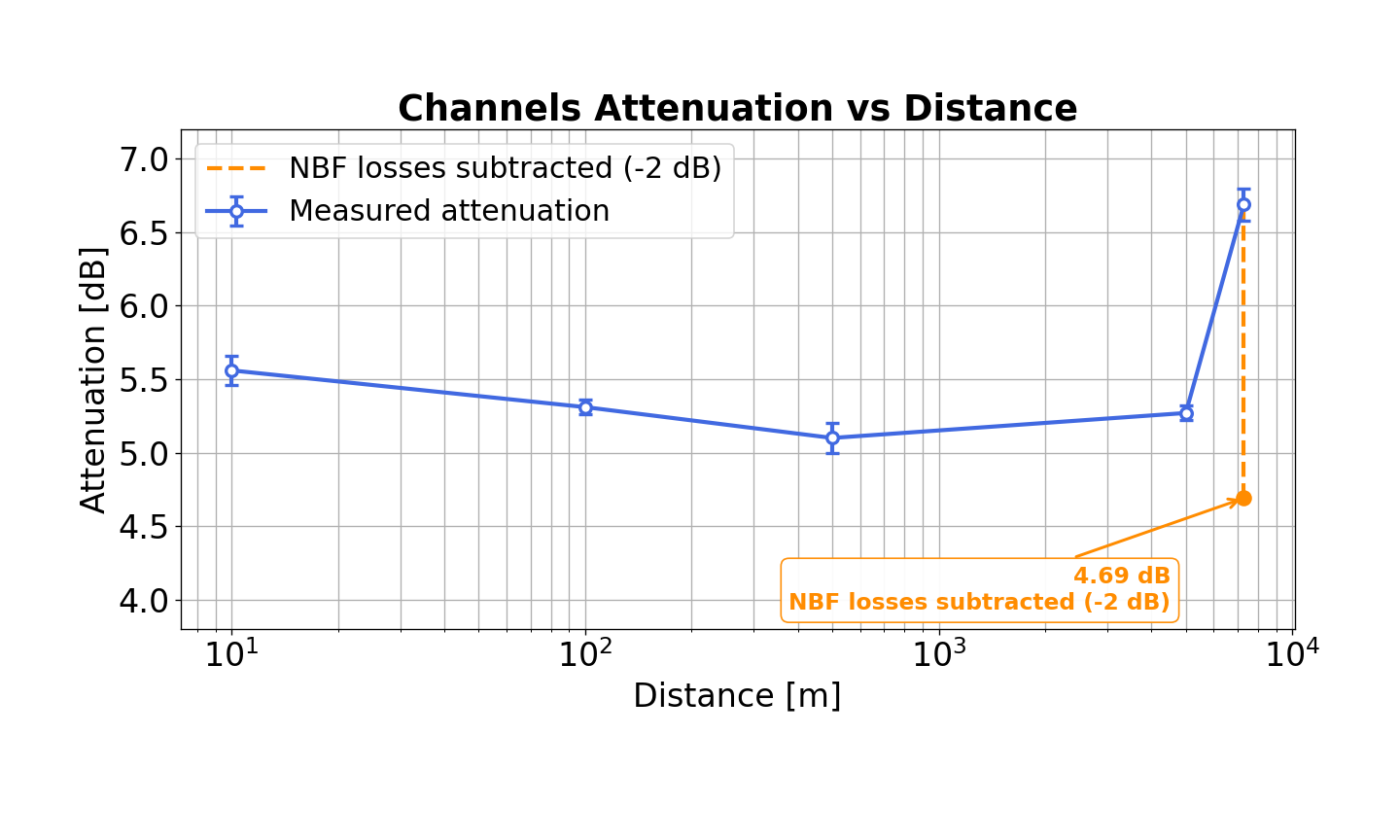}
    \caption{Frequency-response characterization of the quantum channel: relative coincidence attenuation between the single- and double-etalon configurations, as a function of channel length. The attenuation does not grow with propagation length. Instead, it is dominated by the insertion loss of the second $6$~GHz etalon, with the larger value on the $7.3$~km loop consistent with the additional insertion loss of the narrowband filter present only on that link. The absence of a length-dependent trend indicates no measurable propagation-induced spectral shift or distortion. Error bars denote the Poissonian statistical uncertainty.}
    \label{fig:new:15}
    \hrulefill
\end{figure}

The last degree of freedom we consider is frequency. Unlike polarization and timing, the photon's spectral content is not expected to be strongly perturbed by a passive fiber, but on a deployed link this must be verified rather than assumed, since spectral shifts or distortions would compromise frequency-encoded and wavelength-multiplexed schemes.

We probe the channel's spectral response over a representative acquisition,
by comparing the coincidence rate transmitted through a single $6$~GHz etalon with that transmitted through two cascaded, frequency-aligned $6$~GHz etalons placed before and after the channel (Appendix~\ref{app:A.5}). A frequency shift or spectral broadening introduced by propagation would misalign the spectrum of the photon filtered by the first etalon with respect to the second, and reduce the two-filter coincidence rate beyond what the filter's own insertion loss accounts for. Because the measurement is a relative comparison between the one- and two-filter configurations, all losses upstream of the second etalon cancel out and do not bias the result.

Across all channel lengths, the relative attenuation between the one- and two-filter configurations lies in the range $5$--$6.7$~dB (Fig.~\ref{fig:new:15}), dominated by the $\approx 3$~dB insertion loss of the second etalon together with the connector losses of the modified configuration. Crucially, this attenuation does \emph{not} grow monotonically with propagation length: it decreases slightly from the $10$~m to the $500$~m spool and then rises again, reaching its largest value on the $7.3$~km loop. Such a non-monotonic behavior is inconsistent with a propagation-induced spectral distortion, which would instead accumulate with fiber length. The largest value, on the deployed loop, is consistent with the additional insertion loss of the narrowband filter (NBF) -- present only on that link to suppress the Raman and crosstalk background (Sec.~\ref{sec:3}) -- whose peak transmission of about $-2$~dB (Fig.~\ref{fig:app:03}) is by itself sufficient to account for the $\sim 1.4$~dB excess over the spools, leaving no room for a spectral effect of the fiber. The error bars denote the Poissonian statistical uncertainty and account only for photon-counting noise; the residual channel-to-channel variations exceed them and are therefore systematic, of instrumental origin, rather than statistical fluctuations.

Within this picture, no length-dependent, propagation-induced attenuation is observed. We therefore conclude that the channel introduces no measurable frequency shift or spectral distortion of the transmitted photons.

This is a favorable result for network design: unlike polarization and timing, the frequency degree of freedom is essentially stable over deployed metropolitan fiber, requiring no active compensation. Frequency-encoded qubits and wavelength-multiplexed quantum channels can thus be operated over such links without the recalibration that the other two degrees of freedom demand.

\section{QUANTUM-STATE DEGRADATION MODELS}
\label{sec:5}

The characterizations of Sec.~\ref{sec:3} and Sec.~\ref{sec:4} quantify, one impairment at a time, how a deployed fiber degrades a propagating quantum state. To be useful for network engineering, however, these measurements must be distilled into \textit{compact, reusable models}. That are a small set of parameters and closed-form expressions that a designer can evaluate without repeating the experiments, in the same spirit in which classical link budgets rest on simple path-loss models rather than on the full electromagnetic solution~\cite{RapSunMay-13}. Following the criteria that make such models useful -- few parameters, an intuitive basis, closed-form expressions, and stability across datasets -- we distil our measurements into per-link degradation models that a designer can estimate once and reuse to predict performance.

We develop such a model for each of the two degrees of freedom that a deployed metropolitan link degrades appreciably: polarization and timing. Consistent with the discussion of Sec.~\ref{sec:4}, these models are \textit{empirical and representative} rather than universal: their parameters vary between acquisition campaigns and should be read as central estimates with an explicit spread, not as invariant constants of a given fiber length. The frequency degree of freedom, found stable in Sec.~\ref{sec:4.4}, requires no such model and enters a link budget only as a negligible contribution. Together, these models provide the compact ingredients that a quantum link budget requires: given the few parameters of a link, a designer obtains directly the expected fidelity, the induced error.

Before turning to the degradation models, we collect the interference contributions characterized in Sec.~\ref{sec:3} into the summary of Table~\ref{tab:snr_noise_summary}. Expressing them as the fraction of classical power coupled into the quantum channel makes them reusable across links, independently of the specific traffic load. At its C-band minimum near $1535$~nm, the Raman process couples a fraction $\sim 2.7\times10^{-10}$ of the launched O-band power into the receiver bandwidth, while inter-fiber crosstalk couples $\sim 10^{-10}$ of the power carried by neighboring fibers. Remarkably, the two mechanisms -- one from co-propagating traffic in the same fiber, the other from classical channels in adjacent fibers -- inject interference at the same order of magnitude, though with different spectral character: the Raman contribution is broadband and wavelength-dependent (Eq.~\eqref{eq:04}), whereas crosstalk is narrowband and localized at the classical carriers. These fractions, together with the intrinsic-noise floor, are the noise-side input to the link budget.

\begin{table}[t]
    \centering
    \small
    \renewcommand{\arraystretch}{1.2}
    \caption{Interference and noise contributions to the QSINR of a deployed quantum link (Sec.~\ref{sec:3}). Interference terms are expressed as the fraction of classical power coupled into the quantum channel, within the $25$~GHz receiver bandwidth; intrinsic terms as absolute count rates.}
    \label{tab:snr_noise_summary}
    \begin{tabular*}{\columnwidth}{@{\extracolsep{\fill}}lc@{}}
        \hline\hline
        \textbf{Contribution} & \textbf{Level}\\
        \hline
        \hline
        \multicolumn{2}{l}{\textit{Intrinsic noise (count rate)}}\\
            \quad Detector dark counts & $<100$~cps \\
            \quad Ambient background & $\sim 3\times10^{4}$~cps \\
        \hline
        \multicolumn{2}{l}{\textit{Interference (coupled power fraction)}}\\
            \quad Raman scattering (at $1535$~nm min.) & $\sim 2.7\times10^{-10}$ \\
            \quad Inter-fiber crosstalk & $\sim 10^{-10}$ \\
        \hline\hline
    \end{tabular*}
\end{table}

\subsection{POLARIZATION DISTORTION MODEL}
\label{sec:5.1}

The polarization measurements of Sec.~\ref{sec:4.2} showed that the reconstructed state wanders progressively on the Poincar\'e sphere, at a rate that grows with the exposure of the link to its environment. We now distill this behavior into a compact, per-link model of how the polarization fidelity decays with elapsed time.

\begin{figure*}[t]
    \centering
    \includegraphics[width=\linewidth]{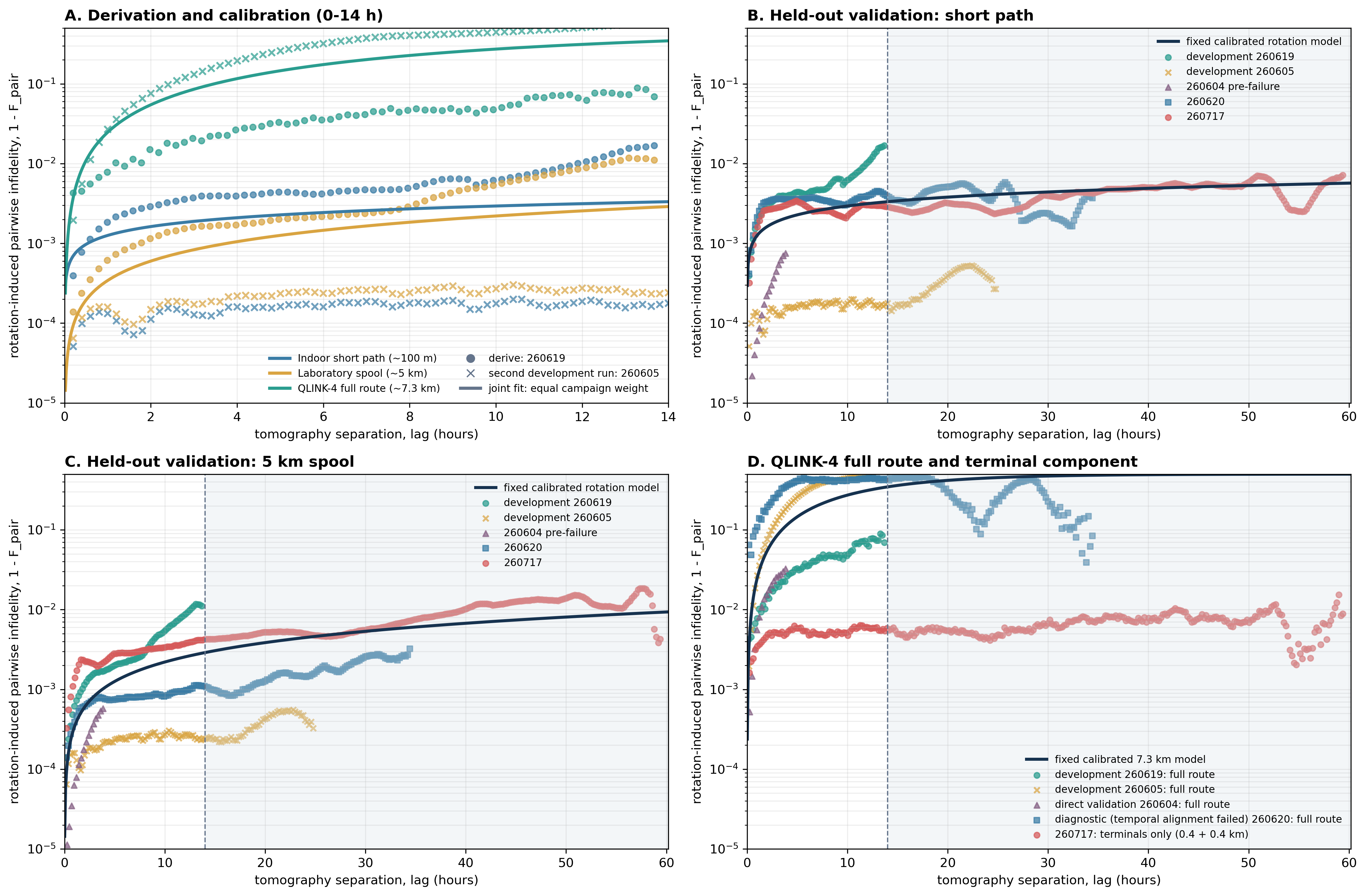}
    \caption{\textbf{Projected-state polarization-drift model.} Projected-state polarization drift versus elapsed time for the three links. Markers denotes the measured pairwise-directional-infidelity $1-F_\mathrm{pair}$. Solid lines are the fitted stretched-exponential model of Eq.~\eqref{eq:pol_stretched}. The drift amplitude grows from the laboratory spools to the deployed $7.3$~km loop, and the exponent is sub-linear on the spools but super-linear ($\alpha>1$) on the deployed loop. Shaded regions indicate extrapolation beyond the 14 h fitting window.}
    \label{fig:pol-model}
    \hrulefill
\end{figure*}

The model is built from coincidence-conditioned tomography, and is therefore a model of the channel's \textit{intrinsic} polarization drift: the coincidence filtering described in Appendix~\ref{app:A.4} suppresses the dark counts and classical-traffic background of Sec.~\ref{sec:3}, so the predicted fidelity is an \textit{upper bound} on what a practical heralding-free receiver achieves, which is further degraded by the noise characterized there.

The observable is directional: for each pair of tomography snapshots separated by a lag~$t$, we form the autocorrelation of the reconstructed Stokes directions,
\begin{equation}
    C(t) = \mathbb{E}\!\left[\mathbf{u}(t')\cdot\mathbf{u}(t'+t)\right], \qquad
    \mathbf{u} = \mathbf{r}/|\mathbf{r}|,
    \label{eq:pol_corr}
\end{equation}
related to the mean pairwise fidelity by $F_\mathrm{pair}(t) = \left[1+C(t)\right]/2$ plotted in Fig.~\ref{fig:new:09}, and we find it well described by a stretched exponential,
\begin{equation}
    C(t) = \exp\!\left[-B\,t^{\alpha}\right].
    \label{eq:pol_stretched}
\end{equation}
Physically, this is a \textit{stochastic rotation} model: the polarization direction diffuses on the Poincar\'e sphere under the time-varying birefringence of the fiber, and $(B,\alpha)$ are the two parameters a designer estimates per link. The amplitude $B$ sets the decorrelation rate, while the exponent $\alpha$ describes how the drift grows with lag: $\alpha \approx 1$ corresponds to ordinary diffusive growth of the angular variance, whereas $\alpha \neq 1$ signals temporally correlated or non-stationary drift.

\begin{figure*}[t]
    \centering
    \includegraphics[width=\linewidth]{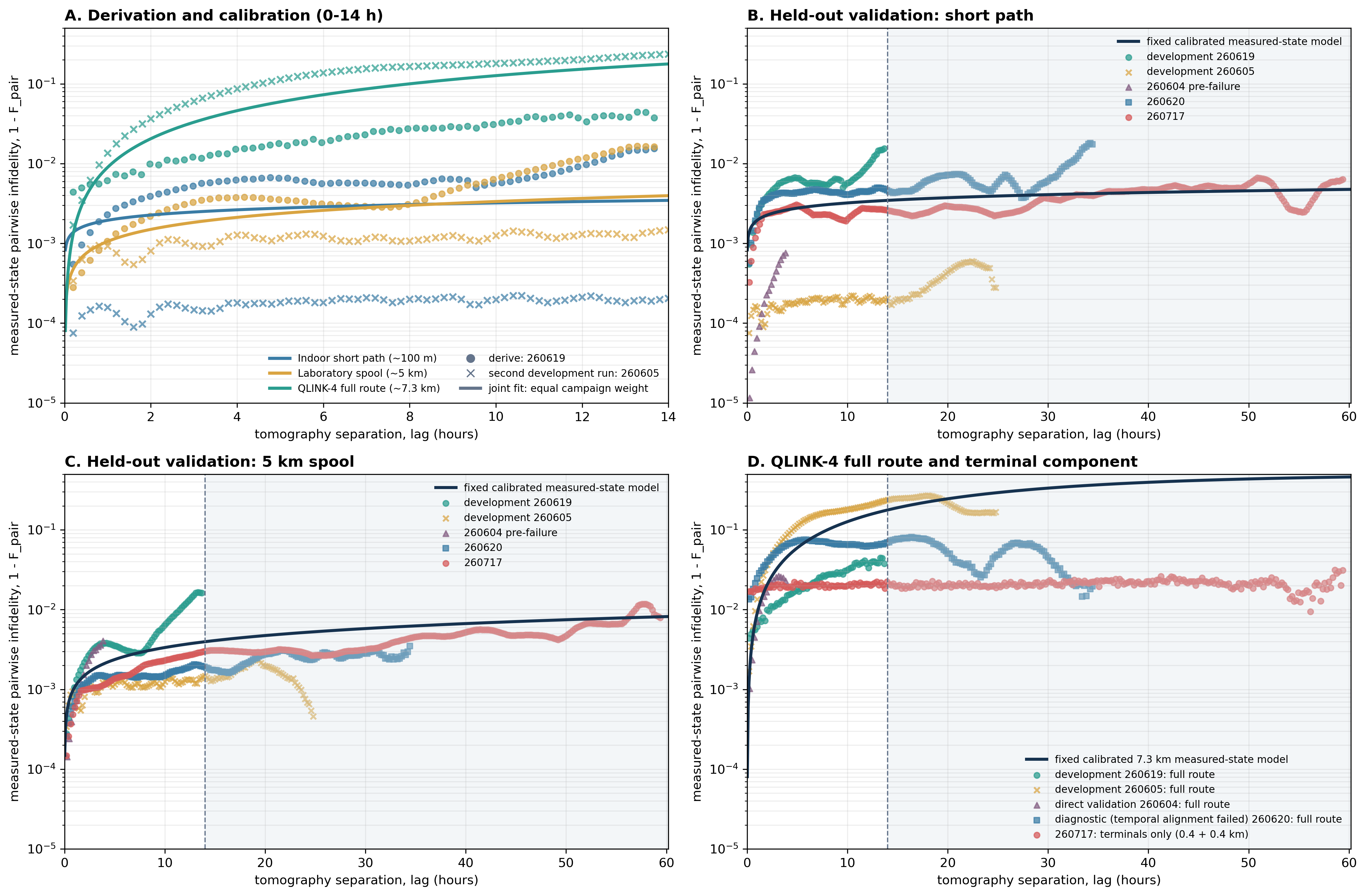}
    \caption{\textbf{Measured-state polarization-drift model.} Measured-state fidelity drift versus elapsed time for the three links. Markers are the measured pairwise infidelity $1-F_\mathrm{pair}$, computed from the mixed-qubit fidelity between the non-projected Bloch vectors (Eq.~\eqref{eq:pol_measured}); solid lines are the fitted stretched-exponential model. Unlike the projected-state model of Fig.~\ref{fig:pol-model}, this retains the reconstructed Bloch-vector radius, so it captures both directional drift and changes in state purity. Shaded regions indicate extrapolation beyond the 14 h fitting window.}
    \label{fig:pol-measured}
    \hrulefill
\end{figure*}

Table~\ref{tab:pol_model} reports $(B,\alpha)$ estimated from coincidence tomography, and Fig.~\ref{fig:pol-model} shows the corresponding fits. The model is built in three stages\footnote{In the format \texttt{YYMMDD\_HHMMSS}: the model is derived on the clean \texttt{260619\_163039} campaign and calibrated by a joint equal-weight fit on \texttt{260619\_163039} and \texttt{260605\_152440}. It is then evaluated on held-out campaigns \texttt{260604\_181001}, \texttt{260620\_075622}, and \texttt{260717\_154348}; for the deployed loop, \texttt{260620\_075622} is treated as diagnostic only (temporal-alignment), run \texttt{260604\_181001} provides a short pre-failure full-route validation, and \texttt{260717\_154348} probes only the in-building terminal component, not the full route.}: the functional form and an initial estimate are derived on one development run; the final parameters are then obtained from a joint fit over two development runs, weighted equally per campaign; and the fixed model is then checked against held-out data -- a short pre-failure full-route acquisition and an in-building terminal-component measurement -- that do not influence the parameters (Fig.~\ref{fig:pol-measured}). Two features stand out. First, the drift amplitude is far larger on the deployed loop than on either laboratory spool, and -- notably -- the $5$~km spool is more stable than the $100$~m indoor path: polarization stability is governed by the environmental exposure of the route, not by its length, so the parameters must be read as route-class specific rather than as a universal fiber-length law. Second, the exponent differs qualitatively between route classes: the laboratory spools drift sub-linearly ($\alpha < 1$, decorrelating quickly then more slowly), whereas the deployed loop drifts \emph{super-linearly} ($\alpha \approx 1.2$), its decorrelation accelerating over the acquisition until the pairwise fidelity falls substantially by the end of the $14$~h window. This super-linear behavior on the deployed link is the least constrained result of the model --- it rests on two development campaigns with limited coincidence statistics on the full route --- and calls for confirmation on a larger dataset.

\begin{table}[t]
    \centering
    \caption{Parameters of the polarization rotation model of Eqs.~\eqref{eq:pol_corr}--\eqref{eq:pol_stretched}, from coincidence tomography, with the predicted pairwise fidelity after one and fourteen hours. $B$ is expressed in units of h$^{-\alpha}$ with the exponent $\alpha$ of each row. Values are representative central estimates; the deployed-loop parameters rest on two development campaigns with limited full-route statistics (see text).}
    \label{tab:pol_model}
    \begin{tabular}{lcccc}
        \hline\hline
        Link & $B$ [h$^{-\alpha}$] & $\alpha$ & $F_\mathrm{pair}(1\,\mathrm{h})$ & $F_\mathrm{pair}(14\,\mathrm{h})$ \\
        \hline
        Indoor $100$~m    & $2.5\times10^{-3}$ & $0.37$ & $0.9987$ & $0.9967$ \\
        Spool $5$~km      & $6.8\times10^{-4}$ & $0.81$ & $0.9997$ & $0.9971$ \\
        Deployed $7.3$~km & $5.1\times10^{-2}$ & $1.19$ & $0.9753$ & $0.6540$ \\
        \hline\hline
    \end{tabular}
\end{table}

\begin{figure*}[t]
    \centering
    \includegraphics[width=\linewidth]{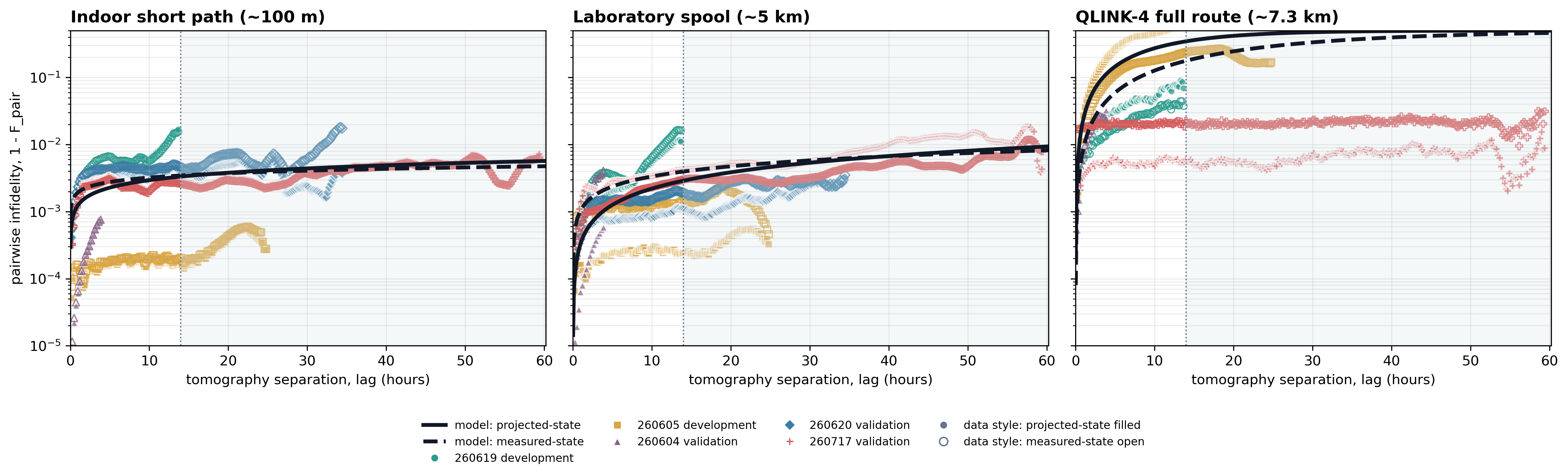}
    \caption{\textbf{Comparison of projected-state and measured-state fidelity decays.} Comparison of the projected-state and measured-state fidelity decays for the three links. The projected-state model retains only the Poincar\'e-sphere direction (Eq.~\eqref{eq:pol_stretched}), shown as solid curves and filled markers, while the measured-state model retains the full Bloch vector, including its radius, through the mixed-qubit fidelity of Eq.~\eqref{eq:pol_measured}, shown as dashed curves and open markers; dataset colors are preserved across campaigns. The close agreement across all routes confirms that the dominant instability is directional drift rather than a change in reconstructed state radius, so the projected-state model is not an artifact of the projection. The deployed-loop points from the \texttt{260717} campaign correspond to the in-building terminal component only, not to the full deployed route. Shaded regions indicate extrapolation beyond the 14 h fitting window.}
    \label{fig:pol-overlay}
    \hrulefill
\end{figure*}

Because the projected-state model retains only the direction of each Stokes vector, it is by construction insensitive to changes in the reconstructed Bloch-vector radius, namely, to apparent purity loss from background, dark counts, or tomography bias as well as any non-unitary effect. To verify that this projection does not drive our conclusions, we repeated the analysis with a companion \textit{measured-state} model that retains the radius and computes the mixed-qubit fidelity between the non-projected vectors,
\begin{equation}
    F(\mathbf{r},\mathbf{s}) = \tfrac{1}{2}\!\left[1 + \mathbf{r}\cdot\mathbf{s} + \sqrt{1-|\mathbf{r}|^2}\sqrt{1-|\mathbf{s}|^2}\,\right],
    \label{eq:pol_measured}
\end{equation}
with $\rho = \tfrac{1}{2}(I+\mathbf{r}\cdot\boldsymbol{\sigma})$ and the Bloch-vector fidelity
\begin{equation}
    F(\mathbf{r},\mathbf{s}) = \tfrac{1}{2}\!\left[1 + \mathbf{r}\cdot\mathbf{s} + \sqrt{1-|\mathbf{r}|^2}\sqrt{1-|\mathbf{s}|^2}\,\right],
    \label{eq:pol_measured}
\end{equation}
and is fitted with the same stretched-exponential form. As shown in the overlay of Fig.~\ref{fig:pol-overlay}, the two models yield comparable fidelity decays across all routes, confirming that the dominant instability in these campaigns is directional drift rather than large changes in reconstructed radius, and that the projected-state model is not an artifact of the projection.

We stress the empirical status of these results. The parameters are representative central estimates with a campaign-to-campaign spread, not universal constants; the deployed-loop values, drawn from two development campaigns with limited full-route statistics, are the most tentative. Evaluated against held-out campaigns without refitting, the model reproduces the indoor and spool behavior well, while the terminal-only deployed measurement is a component-sensitivity check rather than a full-route validation. Within these limits, the model provides exactly what a link budget needs: a compact, measurable law that predicts polarization fidelity as a function of exposure time, and hence the interval at which polarization compensation must intervene.

\subsection{TEMPORAL DRIFT MODEL}
\label{sec:5.2}

\begin{figure*}[t]
    \centering
    \includegraphics[width=\linewidth]{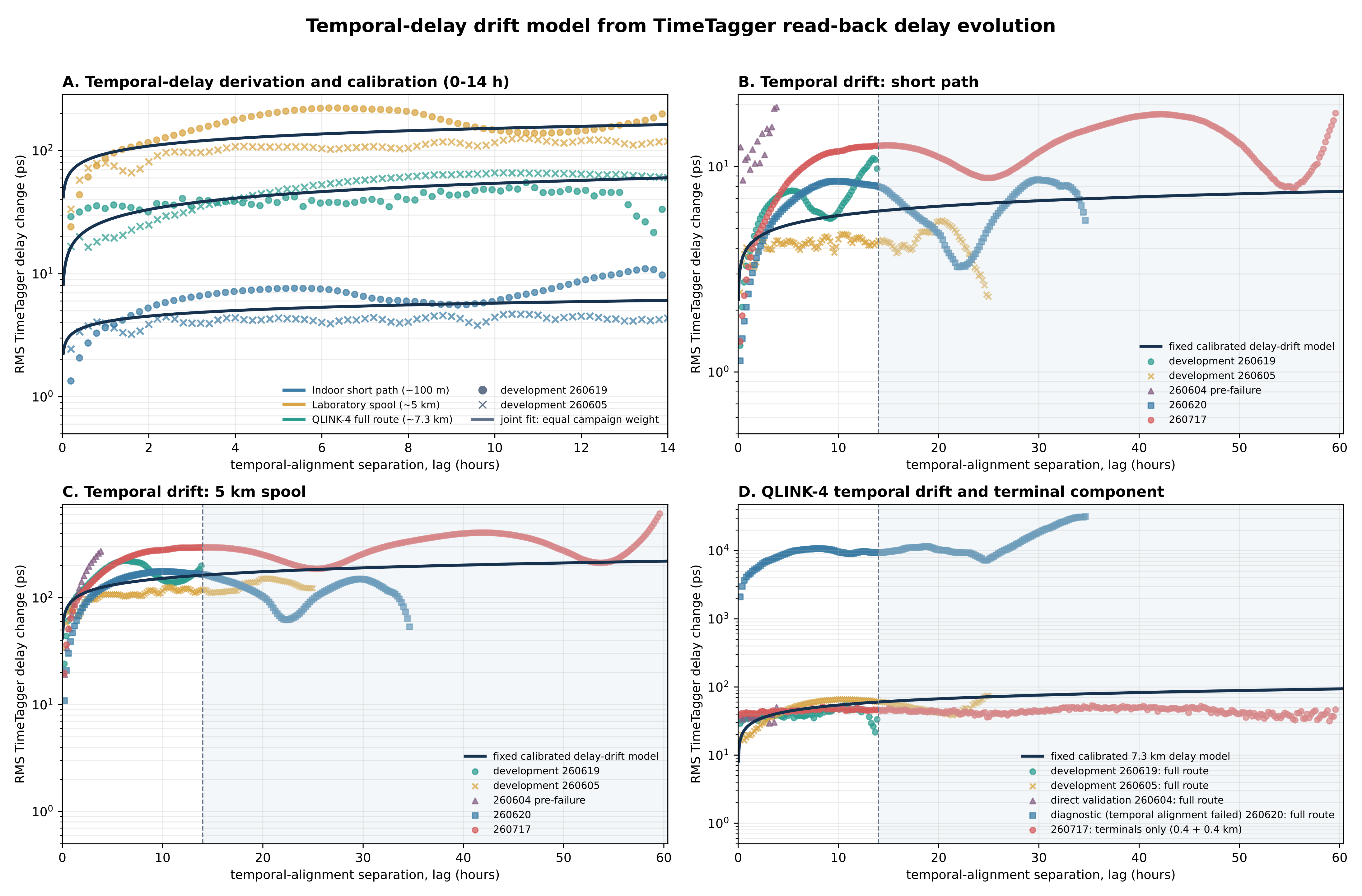}
    \caption{RMS drift of the central delay (temporal-correlation peak position) versus elapsed time for the three links. Markers are the measured pairwise RMS variations of the applied alignment delay; solid lines are the fitted power-law model of Eq.~\eqref{eq:temp_powerlaw}. The $5$~km laboratory spool exhibits the largest drift, while the deployed $7.3$~km loop is markedly more stable, and the sub-linear exponent ($\beta < 1$) indicates a sub-diffusive accumulation.}
    \label{fig:temp-model}
    \hrulefill
\end{figure*}

The temporal characterization of Sec.~\ref{sec:4.3} showed that the propagation delay of a deployed link is not constant but wanders over time (Fig.~\ref{fig:new:11}), so that a synchronization established once is progressively lost. To turn this behavior into a predictive model, we describe how the \emph{central delay} of the link -- the position of the temporal-correlation peak, tracked through the alignment delay applied by the time tagger -- drifts with the elapsed time.

Concretely, we track the applied alignment delay $\tau(t)$ (the fitted peak position of the signal-idler temporal cross-correlation) and form the root-mean-square of its variation between measurements separated by a lag~$t$,
\begin{equation}
    \mathrm{RMS}_{\Delta\tau}(t) = \sqrt{\,\mathbb{E}\!\left[\left(\tau(t'+t)-\tau(t')\right)^2\right]}\,,
    \label{eq:temp_rms}
\end{equation}
which we find is well described\footnote{In the format \texttt{YYMMDD\_HHMMSS}: the model is derived on the clean \texttt{260619\_163039} campaign and calibrated by a joint equal-weight fit on \texttt{260619\_163039} and \texttt{260605\_152440}. It is then evaluated on held-out campaigns \texttt{260604\_181001}, \texttt{260620\_075622}, and \texttt{260717\_154348}; for the deployed loop, \texttt{260620\_075622} is treated as diagnostic only (temporal-alignment failure) and \texttt{260717\_154348} is a terminal-component comparison rather than a full-route validation.} by a power law,
\begin{equation}
    \mathrm{RMS}_{\Delta\tau}(t) = \sigma_{1\mathrm{h}}\,t^{\beta}\,,
    \label{eq:temp_powerlaw}
\end{equation}
where $\sigma_{1\mathrm{h}}$ is the RMS drift of the central delay accumulated after one hour and $\beta$ sets its temporal character.

This model captures the drift of the peak \emph{position} -- the quantity that determines when resynchronization is required -- and is distinct from the peak \emph{width} (the temporal jitter of Fig.~\ref{fig:new:12}), which quantifies the instantaneous temporal indistinguishability and is not modeled here.

Equations~\eqref{eq:temp_rms}--\eqref{eq:temp_powerlaw} constitute the model: given $(\sigma_{1\mathrm{h}},\beta)$, a designer obtains the expected timing drift after any exposure time, and hence the interval after which the accumulated delay exceeds the coincidence window or timing tolerance $\tau_\mathrm{th}$ and resynchronization is required (Eq.~\ref{eq:recal}).

Table~\ref{tab:temp_model} reports $(\sigma_{1\mathrm{h}},\beta)$ estimated for the three links, and Fig.~\ref{fig:temp-model} shows the corresponding fits. Two features stand out, and both reinforce the observations of Sec.~\ref{sec:4.3}. First, and counter-intuitively, the largest drift is not on the longest link: the $5$~km laboratory spool accumulates the most dispersion ($\sigma_{1\mathrm{h}}\approx 95$~ps, exceeding $160$~ps after $14$~h), whereas the deployed $7.3$~km loop is markedly more stable ($\sigma_{1\mathrm{h}}\approx 27$~ps, reaching only $\sim 60$~ps over the same interval). This confirms that temporal stability is governed by the thermal environment rather than by fiber length: the buried, thermally quiet in-ground routing of the deployed loop isolates it from the diurnal temperature swings of the non-climatized laboratory. Second, the exponent $\beta$ is sub-linear ($\beta < 1$) on all links, indicating a sub-diffusive drift that accumulates quickly at first and then more slowly -- so that the resynchronization interval, once the link has settled, is longer than an early extrapolation would suggest.

\begin{table}[t]
    \centering
    \caption{Parameters of the temporal-drift model of Eqs.~\eqref{eq:temp_rms}--\eqref{eq:temp_powerlaw}, with the RMS delay drift predicted after one and fourteen hours of exposure. Values are representative central estimates; see text for the campaign-to-campaign spread.}
    \label{tab:temp_model}
    \begin{tabular}{lcccc}
        \hline\hline
        Link & $\sigma_{1\mathrm{h}}$ [ps] & $\beta$ & RMS$(1\,\mathrm{h})$ [ps] & RMS$(14\,\mathrm{h})$ [ps] \\
        \hline
        Indoor $100$~m    & $4.1$  & $0.15$ & $4.1$  & $6.1$  \\
        Spool $5$~km      & $94.5$ & $0.21$ & $94.5$ & $163$  \\
        Deployed $7.3$~km & $26.9$ & $0.31$ & $26.9$ & $60.2$ \\
        \hline\hline
    \end{tabular}
\end{table}

As for polarization, we stress the empirical status of this model. The observable is the central alignment delay applied by the time tagger, so it inherits the alignment-estimator noise and the discrete read-back quantization of the instrument; the parameters are representative central estimates rather than universal constants, and should not be read as a deterministic propagation-delay law for arbitrary fibers. Evaluated against held-out campaigns without refitting, the model reproduces the indoor and deployed-loop drift within a few picoseconds, while the larger residuals on the $5$~km spool track its intrinsically larger and more variable drift. The deployed-loop series of run \texttt{260620} is excluded from validation, its temporal alignment having failed operationally, and the \texttt{260717} deployed measurement covers only the building-terminal segments and is therefore a component-sensitivity test rather than a full-route validation. Within these limits, the model provides the timing-stability counterpart to the polarization model of Sec.~\ref{sec:5.1}: a compact, measurable law that predicts the resynchronization interval a deployed link demands.

\begin{figure*}[t]
    \centering
    \includegraphics[width=\linewidth]{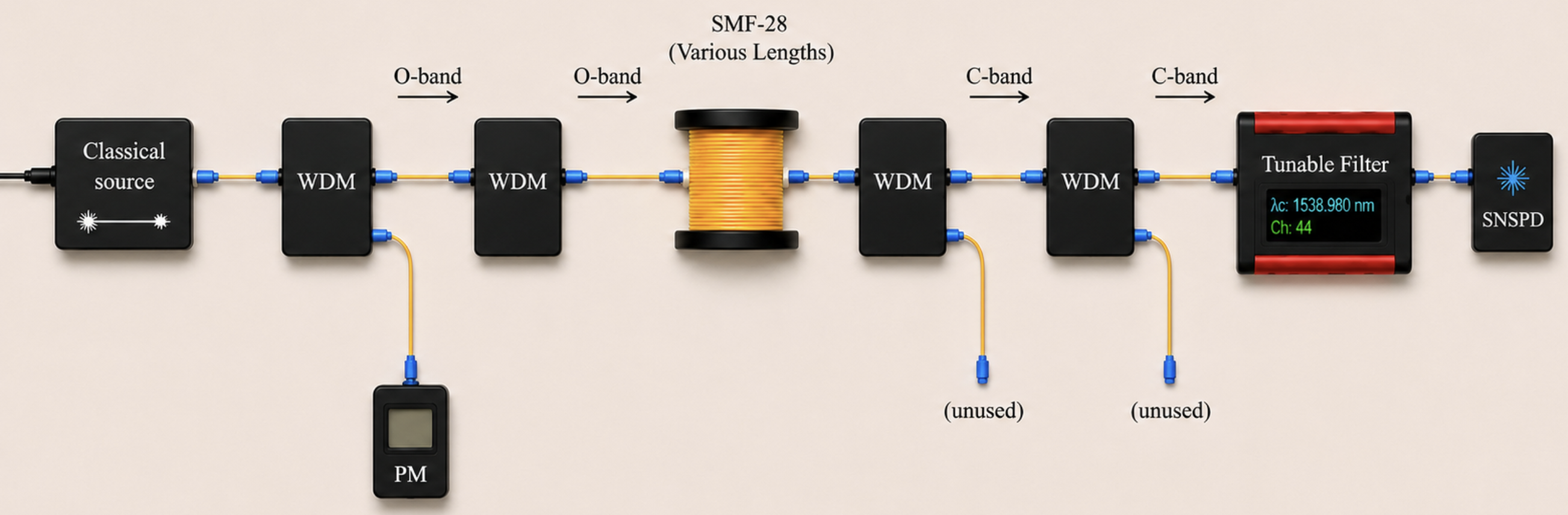}
    \caption{Setup for measuring the C-band SpRS photons generated by an O-band classical signal. A classical source at $1310$~nm is filtered by a cascaded WDM stage to suppress out-of-band components before propagating through the fiber under test. Port~C of the first WDM is connected to a power meter (PM) to monitor source stability. At the receiver, a second WDM cascade isolates the C-band, a narrowband tunable filter scans ITU channels 15 to 63, and the resulting SpRS photons are detected with an SNSPD.}
    \label{fig:app:01}
    \hrulefill
\end{figure*}

\section{DISCUSSION AND CONCLUSION}
\label{sec:6}

We have characterized fiber quantum links spanning two regimes -- controlled laboratory spools and a deployed metropolitan loop -- not as a physics demonstration, but rather an engineering object, described by a small set of measurable parameters organized along two complementary metrics: a photon-counting quantum analog of the SINR for the noise side, and a per-degree-of-freedom analog of the bit-error rate for the state-degradation side.

On the noise side, we quantified the intrinsic and interference contributions directly on the deployed fiber, and developed a predictive Raman model whose C-band minimum near $1535$~nm identifies the least-perturbed operating point in the ITU grid when co-propagating O-band classical signal at $1310$~nm. And we characterized the inter-fiber crosstalk, finding that CWDM C-band classical traffic in neighboring fibers of the same cable couples into the quantum channel at the level of $\simeq 10^{-10}$ of the power carried by the neighboring fibers.

On the degradation side, we distilled the channel-induced drift into two compact per-link models --- a stochastic-rotation model of polarization decorrelation and a power-law model of temporal drift --- finding that polarization and timing require active, quantifiable recalibration while frequency remains essentially stable.

Two aspects of these results deserve emphasis. First, the deployed metropolitan loop does not behave as a scaled-up laboratory fiber: along every degree of freedom, the ordering of the channels is governed by \textit{environmental exposure} rather than by length alone. The $5$~km laboratory spool is the most stable link for polarization, yet exhibits the largest temporal drift, while the buried in-ground loop -- despite being the longest -- is more temporally stable than the shorter spools, plausibly because its thermally quiet underground routing isolates it from the diurnal fluctuations of the laboratory. 

This decoupling of impairment from length is precisely the kind of behavior that only measurements on real infrastructure can reveal, and it cautions against dimensioning a quantum link from its length alone. Second, the two impairment sides are not independent: the classical-traffic interference characterized on the noise side (Sec.~\ref{sec:3}) re-emerges on the degradation side as the residual background that limits the accuracy with which a practical receiver can estimate the transmitted state (Sec.~\ref{sec:5}). A quantum link budget must therefore treat noise and state degradation as related rather than fully independent contributions.

The models developed here are intentionally preliminary and carry explicit limits. They were derived on standard SMF-28 fiber at the metropolitan scale, and their parameters are representative central estimates with an explicit campaign-to-campaign spread rather than universal constants; the deployed-loop parameters, in particular, rest on limited coincidence statistics. Several directions follow naturally. 

On the modeling side, confidence intervals on the fitted parameters, an anisotropic (axis-resolved) extension of the polarization channel, and explicit source-off background acquisitions would sharpen the present estimates. On the scope side, extending the framework to longer intercity and backbone links would test the generality of the two-metric picture. To support reproducibility and the use of these results as engineering references, the complete raw datasets underlying the characterizations of Sec.~\ref{sec:4} will be released, with a permanent identifier, in the camera-ready version of this work.

Together, these results provide the compact, reusable ingredients of a link budget for the Quantum Internet: given a handful of per-link parameters, a network designer can predict the fidelity, the interference-limited signal quality, and the recalibration rates that a deployed quantum link demands -- the same design discipline that underpins classical communication networks, now grounded on measurements over fiber in operational service.

\appendices

\begin{figure*}[!t]
    \centering
    \subfloat[SpRS count rate for different numbers of O-band WDM filters, with two C-band WDM filters fixed at the receiver.\label{fig:app:02.1}]{%
    \includegraphics[width=0.48\textwidth]{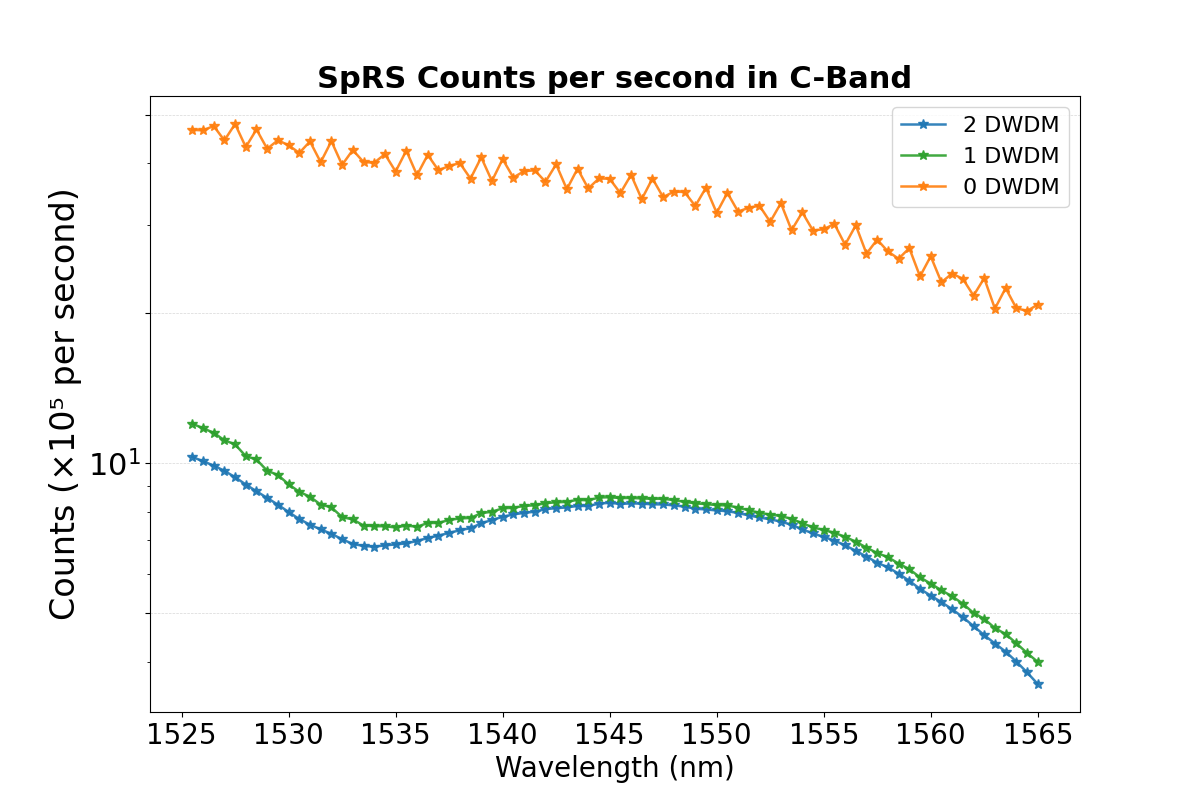}}
    \hfill
    \subfloat[SpRS count rate for different numbers of C-band WDM filters, with two O-band WDM filters fixed at the receiver.\label{fig:app:02.2}]{%
        \includegraphics[width=0.48\textwidth]{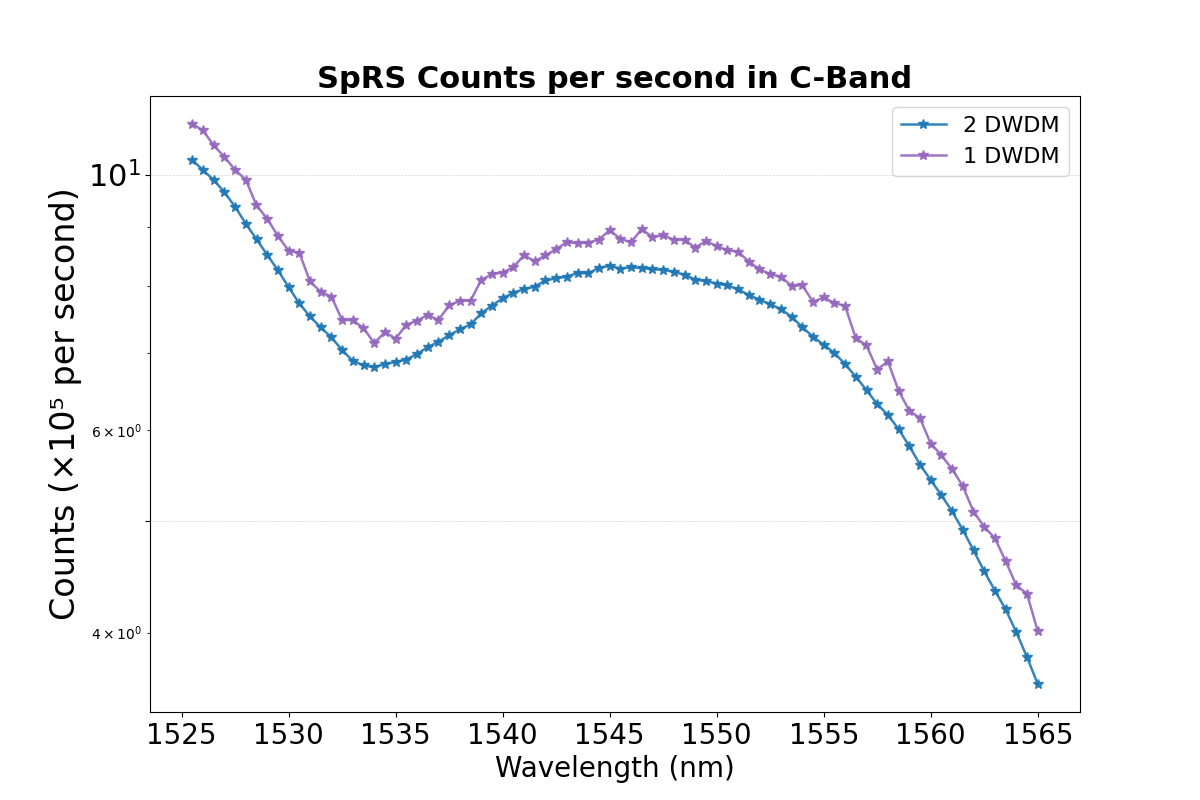}}
    \caption{Measured SpRS count rate $C_{\mathrm{SpRS}}$ for different numbers of cascaded WDM filters. (a) number of O-band filters before transmission. (b) number of C-band filters at the receiver. The double-WDM configuration is adopted in the measurement setup as the best compromise between spectral isolation and insertion loss.}
    \label{fig:app:02}
    \hrulefill
\end{figure*}

\begin{figure}[!t]
    \centering
    \includegraphics[width=\columnwidth]{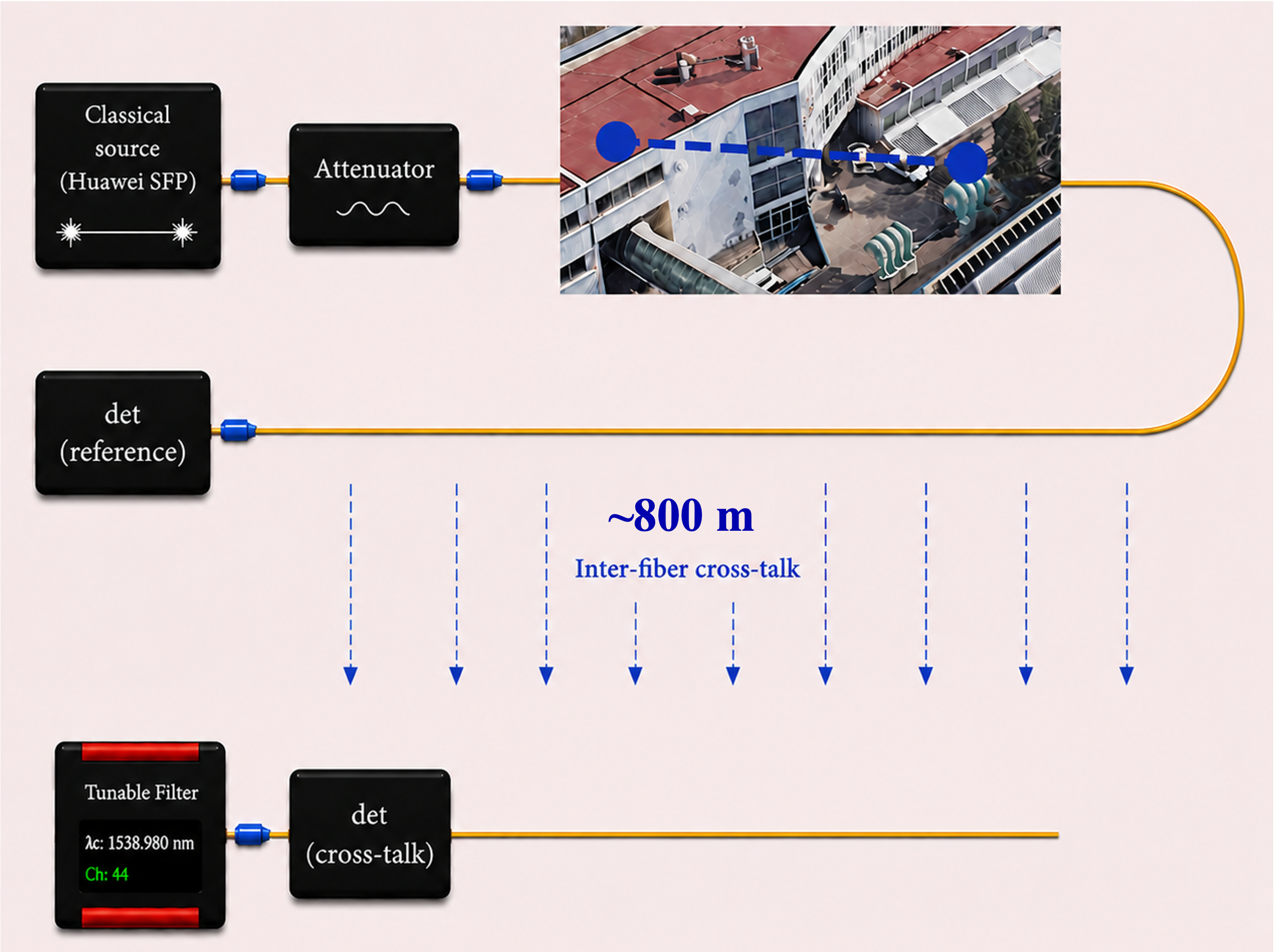}
    \caption{Setup for cross-talk verification: C-band SpRS photon count rate (cps) generated via fiber cross-talk. A Huawei SFP operating at either $1550$~nm or $1570$~nm launches a classical signal into a chain of six serially interconnected fibers (total length $\sim800$~m). The transmitted spectrum is measured after a cascade of optical attenuators and a tunable narrowband filter (NBF). In parallel, an additional fiber of the same cable, carrying no injected signal, is connected to an identical detection stage consisting of a tunable NBF and an SNSPD. Since no optical signal propagates through this fiber, any detected photons originate from coupling with the neighboring active fibers, thereby allowing the spectrum of the induced inter-fiber cross-talk to be reconstructed.}
    \hrulefill
    \label{fig:app:02:bis}
\end{figure}

\section{EXPERIMENTAL SETUPS}
\label{app:A}
This appendix collects the experimental setups used throughout this work.

\subsection{TESTBED INSTRUMENTATION AND METROPOLITAN-SCALE FIBER LOOP}
\label{app:A.1}

\begin{figure*} 
    \centering
    \includegraphics[width=\linewidth]{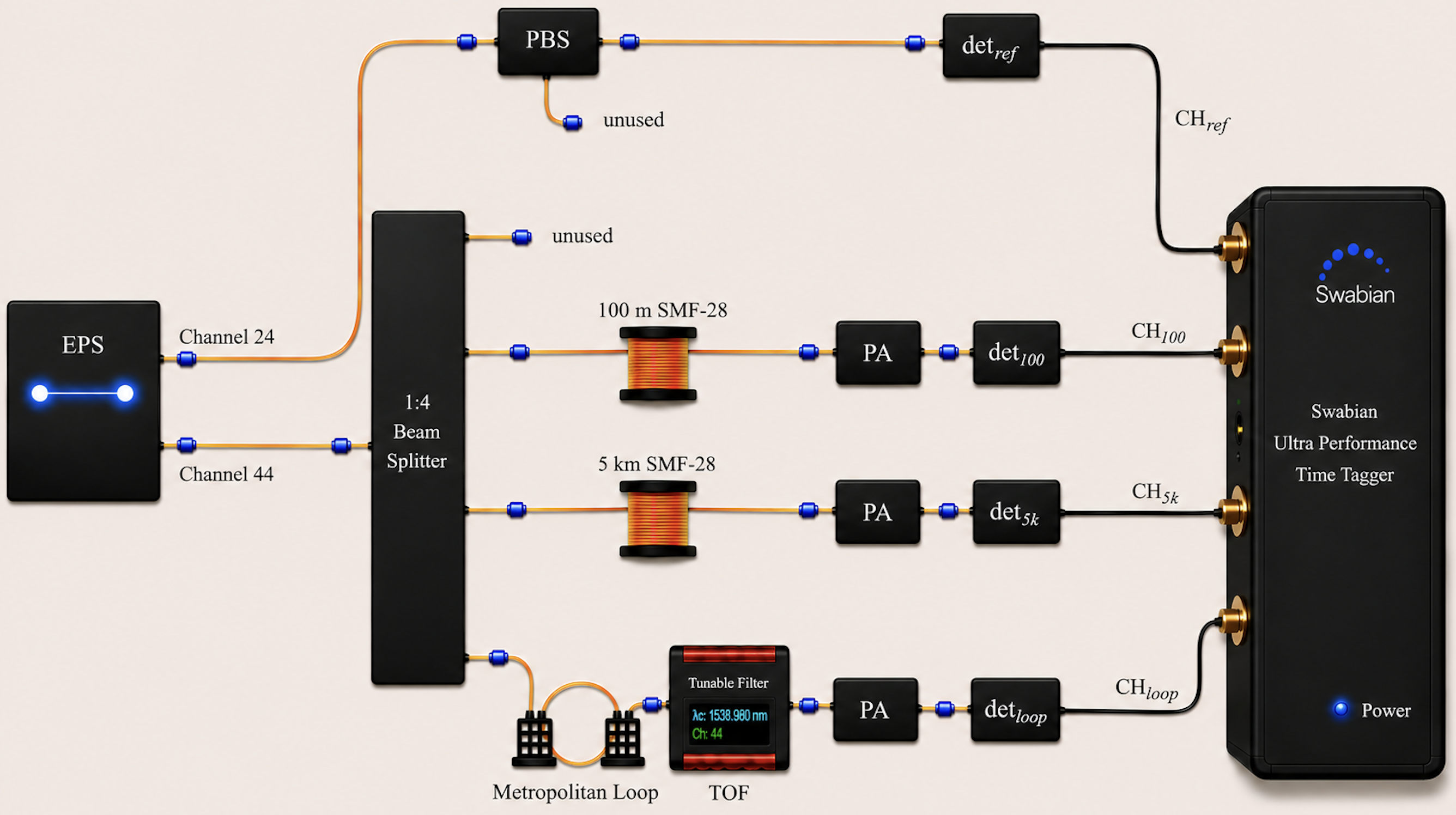}
    \caption{Setup for the temporal and polarization characterizations. The entangled-photon source (EPS) generates photon pairs, which are separated by the internal DWDM filter into ITU DWDM channels 24 (idler) and 44 (signal). The signal photon is divided by a $1{:}4$ beam splitter into the three channels under test: a $100$~m fiber spool ($\mathrm{CH}_{100}$), a $5$~km fiber spool ($\mathrm{CH}_{5\mathrm{k}}$), and the deployed inter-campus loop ($\mathrm{CH}_{\mathrm{loop}}$). Each output is analyzed by a polarization analyzer (PA) and detected by a single-photon detector. On the deployed-loop branch, a narrowband filter (NBF) suppresses out-of-band photons generated by the co-propagating classical traffic. The idler photon is retained on the local reference arm, projected by a polarizing beam splitter (PBS), and detected by the reference detector. The outputs of all four detectors are acquired by a time tagger for temporal-drift measurements and coincidence-based quantum state tomography (QST).}
    \label{fig:app:04}
    \hrulefill
\end{figure*}

The experiments rely on the instrumentation of the national \textit{QuantumInternet.it} testbed and on the \textit{NeQOS} (Networked Quantum Operating System) software platform, which provides automated control, monitoring, synchronization, and data acquisition for the deployed quantum network \cite{Preaparation2} by exploiting the \textit{meta-protocol} paradigm \cite{CacCal-26, CalCac-26, RFCdraft}.

The testbed includes a rack-mounted, plug-and-play Entangled-Photon Source (EPS) generating polarization-entangled photon pairs through an internal Spontaneous Parametric Down-Conversion (SPDC). The generated state is H/V basis is: $\ket{\Phi^+}=\frac{1}{\sqrt{2}}(\ket{H_sH_i}+\ket{V_sV_i})$ where the subscripts $s$ and $i$ refer to the signal and idler, respectively. 
Internal DWDM filtering selects the signal and idler photons on the ITU grid, with the idler  (reference arm) on channel~24 and the signal  (propagating through the quantum link under test) on channel~44. The photons are detected by superconducting-nanowire single-photon detectors (SNSPDs), fabricated and tested by Photon Technology Italy \cite{photec2026} in a rack-mounted configuration. The cryogenic stage operates at $2.2$~K and a pressure of $\sim 10^{-7}$~mbar, and detection events are recorded by a Swabian Ultra Performance time tagger with $8$~ps resolution.


The testbed is equipped with a quantum link deployed within the metropolitan fiber network of the University of Naples Federico II, connecting the Monte Sant'Angelo (MSA) and Engineering Faculty (EF) campuses. The link spans two distinct cables. 
Within the MSA campus, an in-building cable of twelve single-mode fibers connects the laboratory (floor $+2$) to the local Point of Presence (PoP, floor $-1$), which provides access to the metropolitan network. 
Between the campuses, an inter-building metropolitan cable of eight single-mode fibers carries the loop: an outbound path (P1) brings the quantum signal and the co-propagating classical traffic from MSA to EF, and a return path (P2) brings them back, the two being interconnected at EF through a patch fiber that closes the loop. Including both the in-building and inter-building segments, the full round trip amounts to approximately $7.3$~km. The MSA campus hosts the laboratory, where photons are generated and detected after completing the loop.

\subsection{RAMAN INTERFERENCE SETUP}
\label{app:A.2}

The setup used to characterize the C-band SpRS noise generated by O-band classical traffic is shown in Fig.~\ref{fig:app:01}. The classical sources -- a narrow-linewidth laser, a commercial Finisar SFP, and a third Huawei SFP used as an independent validation source -- emit at $1310$~nm at different launch powers. The output is filtered through a cascade of two wavelength-division multiplexing (WDM)\footnote{WDM devices are passive wavelength-selective filters that combine or split optical channels at different wavelengths.} stages to isolate the O-band and suppress any spectral component outside $1310$~nm. Port~C of the first WDM is connected to a power meter (PM) to monitor the source output power in real time, ensuring no significant power fluctuation occurs during acquisition. The O-band signal then propagates through the fiber under test, where it generates the Raman-scattered photons, and is filtered by a second WDM cascade that isolates the C-band. The number of cascaded WDM stages was also investigated: as shown in Fig.~\ref{fig:app:02}, adding a second O-band WDM substantially reduces residual out-of-band counts, while a second C-band WDM smooths the measured spectrum. The double-cascade configuration on both sides of the fiber was therefore adopted as the best compromise between isolation and insertion loss. The C-band output is finally filtered by a tunable narrowband filter\footnote{The tunable narrowband filter exhibits a sharp roll-off at the passband edges, with out-of-band suppression exceeding $-40$~dB, ensuring high channel selectivity across the scanned ITU range.} (NBF) of $25$~GHz bandwidth ($200$~pm), scanned in $500$~pm steps across ITU channels 15 to 63, and detected by an SNSPD. The classical sources and fiber lengths used to build and validate the model are listed in Table~\ref{tab:new:07}.

\begin{figure*}[t]
    \centering
    \includegraphics[width=\textwidth]{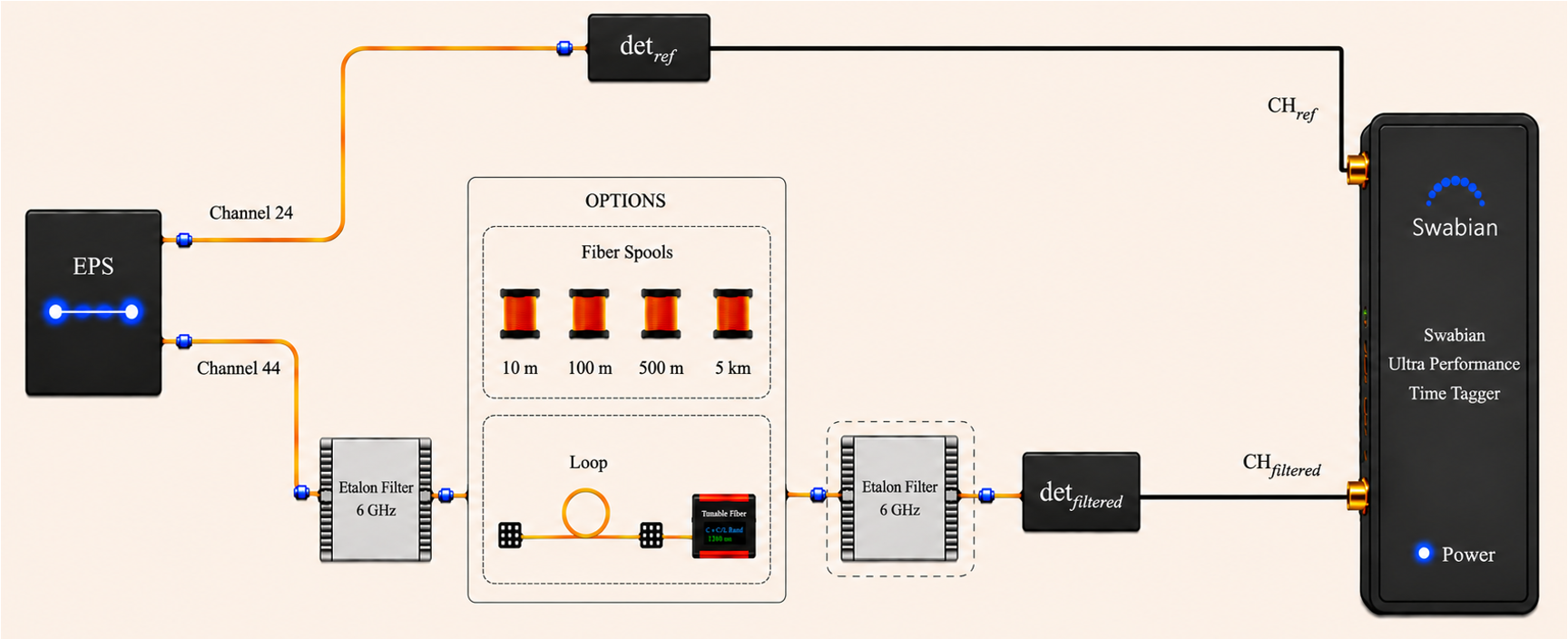}
    \caption{Setup for the frequency-degradation characterization. The idler photon (ITU channel~24) is projected by a polarizing beam splitter (PBS) and detected as a heralding reference. The signal photon (channel~44) is spectrally selected by a $6$~GHz etalon placed upstream of the channel under test -- a $10$~m, $100$~m, $500$~m, or $5$~km spool, or the $7.3$~km deployed loop -- and detected in < with the idler. In a second set of measurements, a further $6$~GHz etalon is introduced at the channel output to probe propagation-induced spectral shifts or distortions. On the deployed loop, a narrowband filter (NBF) centered on channel~44 rejects the out-of-channel Raman and crosstalk background. }
    \hrulefill
    \label{fig:app:05}
\end{figure*}

\subsection{CROSSTALK INTERFERENCE SETUP}
\label{app:A.3}

To verify that the spectral peaks observed on the deployed inter-campus loop (Sec.~\ref{sec:3.4}) originate from inter-fiber coupling, we reproduced the effect in a controlled configuration by exploiting the in-building cable connecting the laboratory (floor $+2$) to the loop's point of presence (floor $-1$) as shown in Fig.~\ref{fig:app:02:bis}. Of the twelve single-mode fibers in this cable, six are patched in series, alternating \textit{down} and \textit{up} spans along the $+2 \leftrightarrow -1$ routing, so that an injected signal propagates over a total length of about $800$~m before returning to the laboratory. A CWDM signal at either $1550$~nm or $1570$~nm is launched into this six-fiber chain. The spectral shape of the transmitted signal is first verified on the active chain to confirm that it is centered at the nominal wavelength. The measurement is then repeated on a separate fiber of the same cable, connected to the detection stage. Since no signal is launched into this fiber, any detected counts can only arise from coupling with the neighboring active fibers within the shared jacket.

\subsection{POLARIZATION AND TEMPORAL DEGRADATION SETUP}
\label{app:A.4}

The polarization and temporal characterizations of Sec.~\ref{sec:4} share a common configuration, designed to probe the three channels simultaneously as shown in Fig.~\ref{fig:app:04}. 

From the EPS, ITU DWDM channels 24 and 44 are selected for idler and signal, respectively, as this channel pair is only marginally affected by the crosstalk generated by the metropolitan classical traffic discussed in Sec.~\ref{sec:3.4}.

The signal photon emerging from the source is divided by a $1{:}4$ beam splitter, three of whose outputs are connected to the channels under test -- the $100$~m spool ($\mathrm{CH}_{100}$), the $5$~km spool ($\mathrm{CH}_{5\mathrm{k}}$), and the deployed inter-campus loop ($\mathrm{CH}_\mathrm{loop}$) -- while the fourth is left unused, so that all three quantum links are measured concurrently under identical source and environmental conditions. The idler photon is retained on the local reference arm and projected by a polarizing beam splitter (PBS) onto a single output port. Conditioning on this projection heralds the signal photon into a well-defined, separable polarization state\footnote{Projecting one photon of the $\ket{\Phi^+}$ pair collapses the other into the corresponding pure state; detecting the idler on a single PBS port therefore prepares a fixed input polarization for the signal.}, which is the state whose channel-induced evolution is reconstructed. Coincidence with this heralding event provides the reference for the tomography. On the deployed inter-campus loop arm, a narrowband tunable filter (NBF) (Appendix.~\ref{app:A.2}) centered on ITU channel 44 rejects the out-of-band photons generated by the classical traffic before detection. Without this filter, the broadband sensitivity of the SNSPD would result in the detection of both DWDM crosstalk and spontaneous Raman-scattering photons over a wide spectral range, rather than only photons within DWDM channel 44.

For the polarization characterization, each beam-splitter arm (corresponding to a different quantum link) is analyzed by a polarization analyzer (PA) -- consisting of a half-wave plate (HWP), a quarter-wave plate (QWP), a second half-wave plate (HWP), and a linear polarizer -- allowing the projective measurements required for QST. For each analyzer setting, two independent acquisitions are performed as motivated in Sec.~\ref{sec:4.1}. One -- referred to in the following as \textit{coincidences} -- records the coincidence counts between the signal detector and the idler reference, from which the polarization state is reconstructed by QST. The other -- referred to in the following as \textit{singles} -- records only the signal-detector singles, without conditioning on the idler, allowing a direct assessment of the impact of background photons and classical-traffic interference. 

\begin{figure}[t]
    \centering
    \includegraphics[width=0.9\columnwidth]{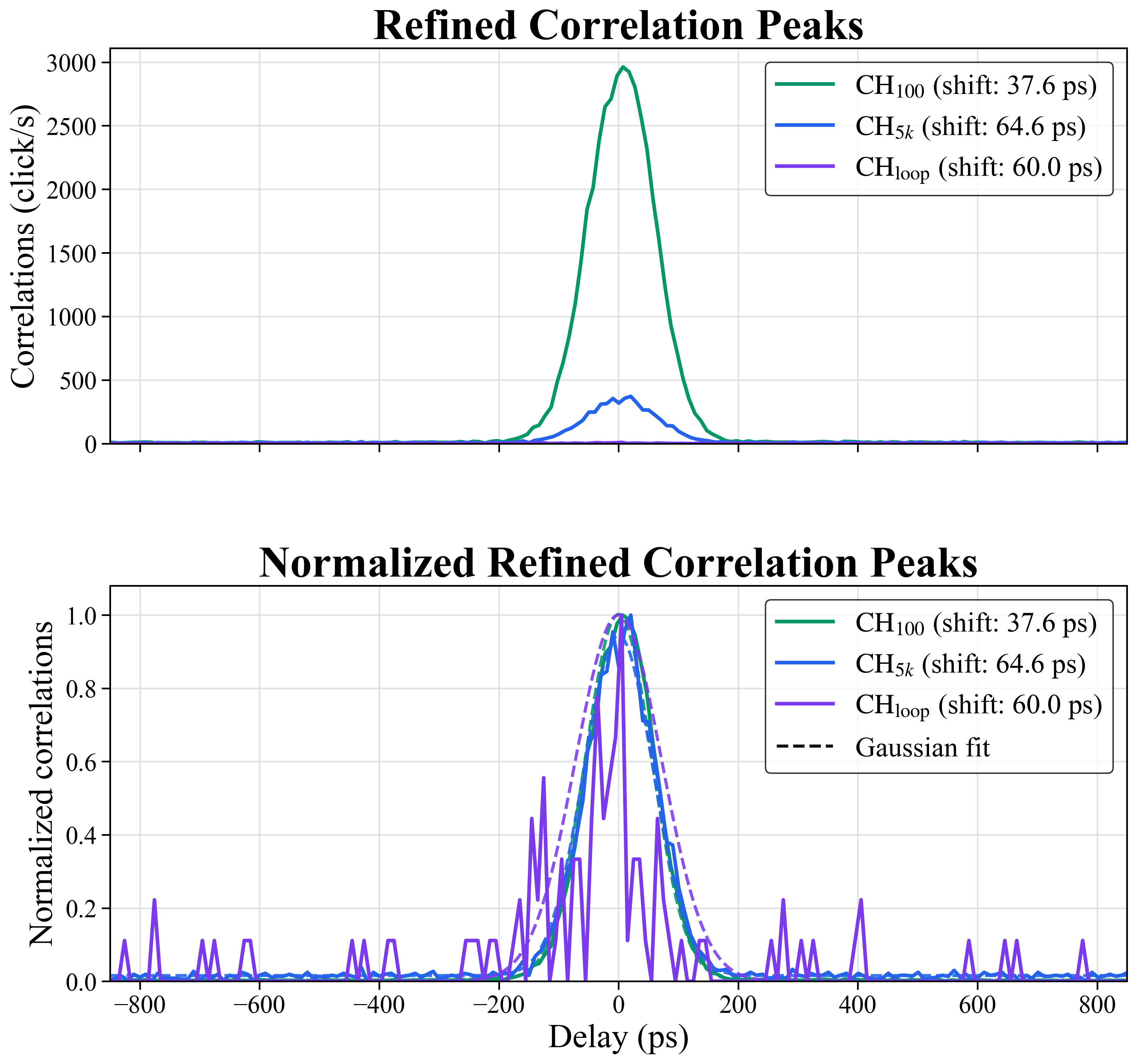}
    \caption{Temporal cross-correlation histograms between the reference (idler) channel and the three channels under test, shifted so that their peaks align at zero delay. The temporal shift $\tau$ applied to each channel is reported in the legend.}
    \hrulefill
    \label{fig:app:04.bis}
\end{figure}

For the temporal characterization, the arrival-time difference between each channel and the reference is obtained from their temporal cross-correlation. Taking the reference (idler) clicks as start events and the channel clicks as stop events, the time tagger accumulates their time differences into a histogram, building the coincidence distribution within a defined window; a pronounced peak marks an excess of correlated counts over the background, and a Gaussian fit to the peak yields, through its mean, the delay of that channel. The resulting temporal cross-correlation histograms, shifted to a common reference time, are shown in Fig.~\ref{fig:app:04.bis} together with the applied delays $\tau$. The histograms of $\mathrm{CH}_{100}$ and $\mathrm{CH}_{5\mathrm{k}}$ are well described by Gaussian profiles, yielding accurate peak positions. In contrast, the correlation function of $\mathrm{CH}_{\mathrm{loop}}$ exhibits a more irregular shape, yet a well-defined peak remains and the fit still determines the delay reliably.

These Gaussian-fitted mean estimates are used not only to quantify the temporal drift discussed in Sec.~\ref{sec:4.3}, but also to compensate the relative timing of the three arms before the coincidence-based polarization measurements. Although not required for the temporal characterization, the polarization analyzers remain in place, since both temporal and polarization measurements are performed within the same experimental campaign using an unchanged optical setup.  As for the polarization compensation, after compensating the relative delays between idler and signal channels, coincidence events are identified using a $1$~ns window. This value is one order of magnitude smaller than the approximately $10$~ns pump repetition period, preventing detections originating from consecutive pump pulses from being associated with the same photon pair.

The Gaussian-fitted mean estimates are used not only to quantify the temporal drift discussed in Sec.~\ref{sec:4.3}, but also to compensate for the relative delays between the idler and the three signal channels before the coincidence-based polarization measurements. Although not required for the temporal characterization itself, the polarization analyzers remain in place because the temporal and polarization measurements are performed within the same experimental campaign without modifying the optical setup. Besides preventing detections originating from adjacent pump pulses from being associated with the same photon pair, this narrow window also suppresses accidental coincidences. For the count rates measured in our experiment, the probability of an accidental coincidence is on the order of $10^{-4}$, making its contribution to the tomographic reconstruction negligible.

\subsection{FREQUENCY DEGRADATION SETUP}
\label{app:A.5}

The frequency-degradation characterization of Sec.~\ref{sec:4.4} uses the configuration of Fig.~\ref{fig:app:05} on the same quantum links employed for the polarization and temporal measurements -- the $100$~m and $5$~km spools and the deployed $7.3$~km loop -- supplemented by $10$~m and $500$~m spools. As in those characterizations, the idler photon (ITU channel~24) is retained on the local reference arm, projected by a polarizing beam splitter (PBS) and detected as a heralding reference, so that all measurements are performed in coincidence. The signal photon (ITU channel~44) is first spectrally selected by an etalon filter\footnote{An etalon is a Fabry--P\'erot filter whose narrow, periodic transmission peaks define a high-resolution spectral reference.} of FWHM bandwidth as low as $6$~GHz placed upstream of the channel, which, combined with the narrowband filter (NBF) centered on channel~44, rejects the Raman scattering and inter-fiber crosstalk of Sec.~\ref{sec:3} on the inter-campus loop.

The spectral response is then probed in two configurations. In the first, the signal photon propagates through the channel under test and is detected in coincidence with the idler, yielding the reference photon rate for each length. In the second, a further etalon filter, also tuned to $6$~GHz and frequency-aligned with the first\footnote{The two etalons have been characterized beforehand and verified to be frequency-aligned. In the absence of any channel, inserting the second etalon reduces the coincidence rate by its nominal insertion loss of about $3$~dB, with no additional distortion.}, is introduced at the channel output before detection. A propagation-induced frequency shift or spectral broadening of the signal would misalign its spectrum with respect to this second etalon and reduce the coincidence rate beyond the filter's own insertion loss. Because the two configurations differ only by the second etalon, all losses upstream of it cancel in the comparison.

\begin{table}[t]
    \centering
    \small
    \renewcommand{\arraystretch}{1.08}
    \caption{Mean absolute relative error between measured and predicted Raman spectra. The $500$~m and $5$~km measurements (narrow-linewidth laser, Finisar SFP) are used for extraction; the Huawei SFP and the $100$~m spool are validation conditions.}
    \begin{tabular}{lcc}
        \hline\hline
        \textbf{Source} & \textbf{Length} & \textbf{$\epsilon$ (\%)}\\
        \hline
        \multirow{3}{*}{Narrow-linewidth laser}
            & $100$~m & 14.9\\ & $500$~m & 5.1\\ & $5$~km & 3.7\\
        \hline
        \multirow{3}{*}{Finisar FTLF1321P1BTL}
            & $100$~m & 21.2\\ & $500$~m & 5.1\\ & $5$~km & 13.3\\
        \hline
        \multirow{3}{*}{Huawei SO-SFP-10GE-LR}
            & $100$~m & 32.2\\ & $500$~m & 9.1\\ & $5$~km & 2.7\\
        \hline\hline
    \end{tabular}
    \label{tab:new:07}
\end{table}

\begin{figure}[t]
    \centering
    \subfloat[\label{fig:cross1}]{%
        \includegraphics[width=0.48\textwidth]{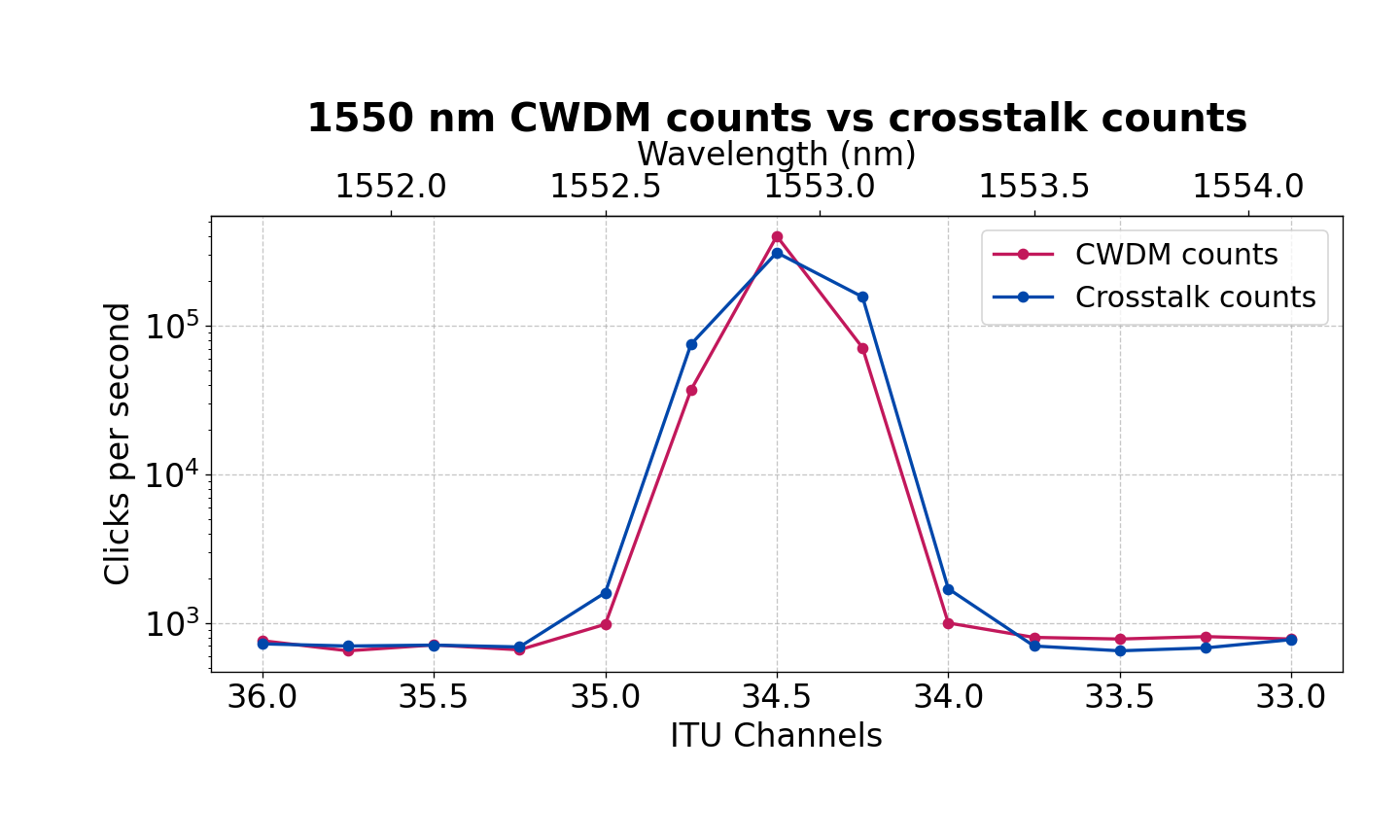}}
    \hfill
    \subfloat[\label{fig:cross2}]{%
        \includegraphics[width=0.48\textwidth]{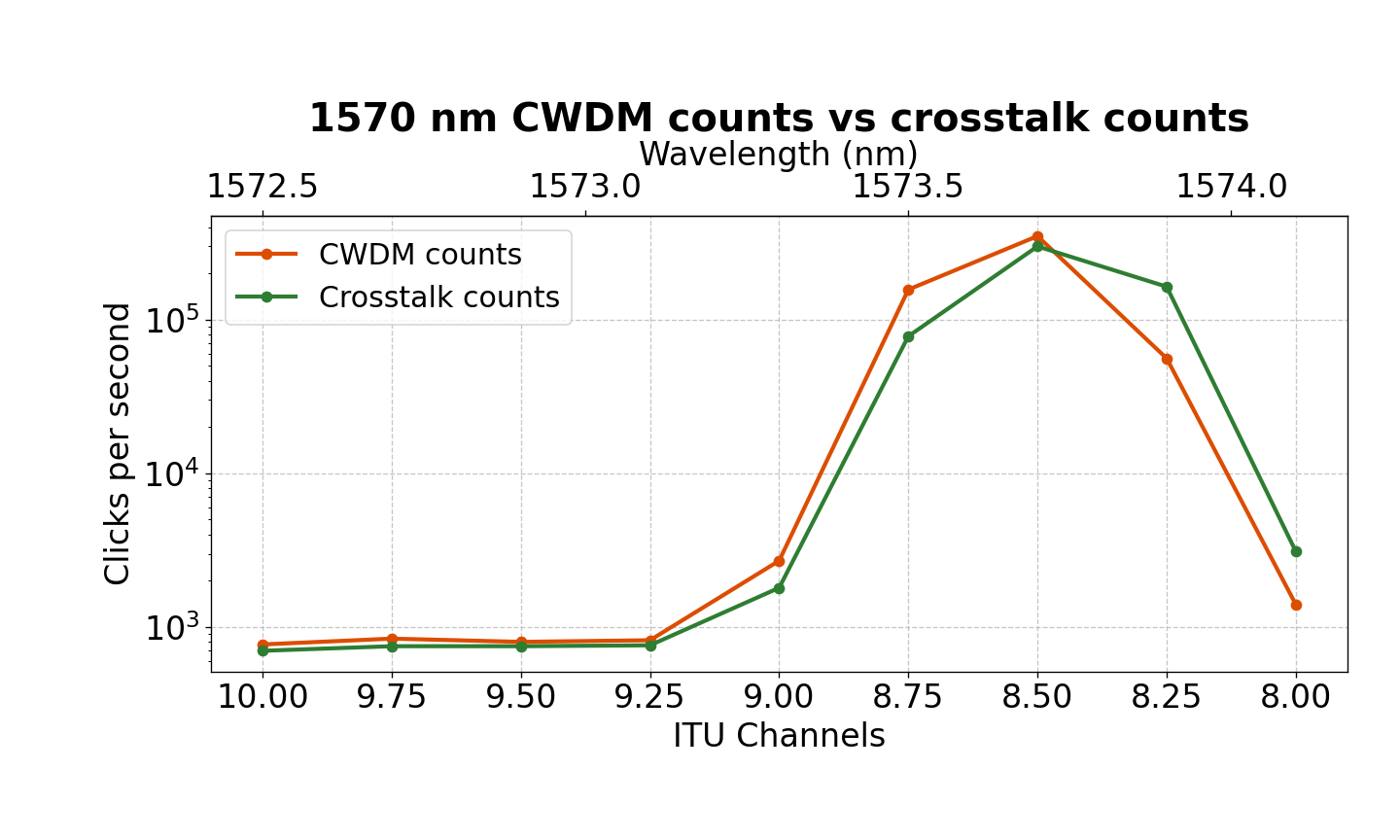}}
    \caption{Crosstalk verification experiment: photon count rate (cps) as a function of wavelength in the C-band for (a) CWDM signal centered at $1550$~nm and (b) CWDM signal centered at $1570$~nm. A CWDM signal is launched into six series-connected fibers of the in-building cable ($\sim 800$~m total), and the spectrum is measured on a separate un-fed fiber of the same cable. In each panel, the counts measured on the active chain (red curve) have been rescaled to the crosstalk level for visual comparison: the actual signal is 10 orders of magnitude higher. The spectral shape detected on the unfed pair closely reproduces that of the active signal, confirming inter-fiber coupling within the shared jacket.}
    \hrulefill
    \label{fig:app:03}
\end{figure}

\section{RAMAN SCATTERING MODEL: EXTRACTION AND VALIDATION}
\label{app:B}

This appendix details the extraction and validation of the Raman-noise model of Sec.~\ref{sec:3.3}; the setup is described in Appendix~\ref{app:A.2}.

The C-band SpRS count rate $C_\mathrm{SpRS}(\lambda_c)$ is measured for an O-band source at $\lambda_o=1310$~nm through a $\Delta\lambda=25$~GHz tunable filter, so the click rate is already integrated over the receiver bandwidth of Eq.~\eqref{eq:04}. To make it source- and length-independent, it is normalized by the launched O-band power $P_o$ ($\simeq-4.2$~dBm for the narrow-linewidth laser, $\simeq-1.4$~dBm for the Finisar SFP) and by the effective interaction length $L_\mathrm{eff}$ (with $\alpha_o=0.35$ and $\alpha_c=0.2$~dB/km), yielding the effective coefficient $\beta_\mathrm{eff}$ of Eq.~\eqref{eq:05}. A low-order polynomial fit of $\beta_\mathrm{eff}$ over $[1525,1565]$~nm already reproduces the measured shape, including the $1535$~nm minimum, and motivates the physically grounded two-mode representation of Sec.~\ref{sec:3.3}, in which Hollenbeck--Cantrell modes~12 and~13 are re-fitted to the data.

The model is validated against conditions excluded from its extraction. Agreement is quantified by the mean absolute relative error
\begin{equation}
    \epsilon(\%) = \frac{100}{N} \sum_{i=1}^{N}
        \frac{\left| C_\mathrm{SpRS}(\lambda_i) - C_\mathrm{model}(\lambda_i) \right|}{C_\mathrm{SpRS}(\lambda_i)},
    \label{eq:app:relative-error}
\end{equation}
reported in Table~\ref{tab:new:07}. A Huawei SO-SFP-10GE-LR transceiver (a different transmitter architecture) and an additional $100$~m spool both agree with the prediction; the larger error at $100$~m follows from the weaker Raman signal, not from a model limitation, as the spectral shape and minimum location are still reproduced. Finally, injecting the O-band signal into the deployed loop reproduces the same normalized C-band spectrum as the laboratory reference, confirming that the model holds over deployed metropolitan fiber.

\section{CROSSTALK INTERFERENCE VALIDATION}
\label{app:C}

Using the setup described in Appendix~\ref{app:A.3}, the induced inter-fiber crosstalk spectrum is reconstructed on the unfed fiber. The spectrum measured on the unfed fiber closely matches that of the active transmission channel for both launch wavelengths (Fig.~\ref{fig:app:03}), despite the absence of any injected optical signal. This agreement confirms that the detected photons originate from inter-fiber coupling, thereby validating the attribution of the deployed-loop spectral peaks to the crosstalk term $N_\mathrm{xt}$.

%

\begin{IEEEbiography}[{\includegraphics[width=1in,height=1.25in,clip,keepaspectratio]{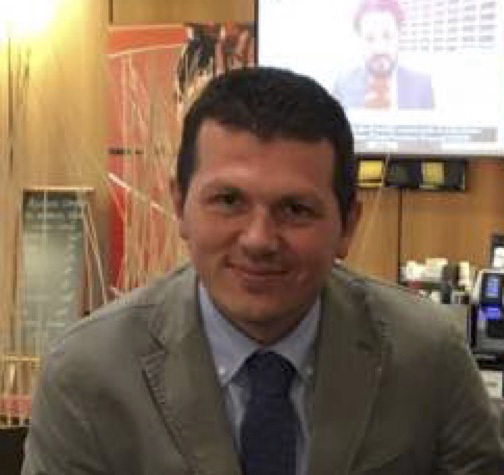}}]{Marcello Caleffi} (\textit{Senior Member, IEEE}) is currently
Professor of advanced quantum networks with the Department of Electrical Engineering and Information Technologies, University of Naples Federico II, Naples, Italy, where he co-founded Quantum Internet
Research Group. His research has appeared in several premier IEEE Transactions and journals. He was the recipient of multiple awards, including the 2024 IEEE Communications Society Award for Advances in Communication and 2022 IEEE Communications Society Best Tutorial Paper Award. He serves as Area Editor of IEEE OPEN JOURNAL OF THE COMMUNICATIONS SOCIETY, and as Editor of IEEE JOURNAL ON SELECTED AREAS IN COMMUNICATIONS, IEEE TRANSACTIONS ON COMMUNICATIONS, IEEE TRANSACTIONS ON QUANTUM ENGINEERING, and IEEE INTERNET COMPUTING. He has served as the Chair and TPC Chair of several premier IEEE conferences. In 2017, he was appointed as the Distinguished Visitor Speaker by the IEEE Computer Society and elected Treasurer of IEEE ComSoc/VT Italy Chapter. He was appointed as a member
of IEEE New Initiatives Committee by the IEEE Board of Directors in 2019 and as an IEEE ComSoc Distinguished Lecturer in 2023.
\end{IEEEbiography}

\begin{IEEEbiography}[{\includegraphics[width=1in,height=1.25in,clip,keepaspectratio]{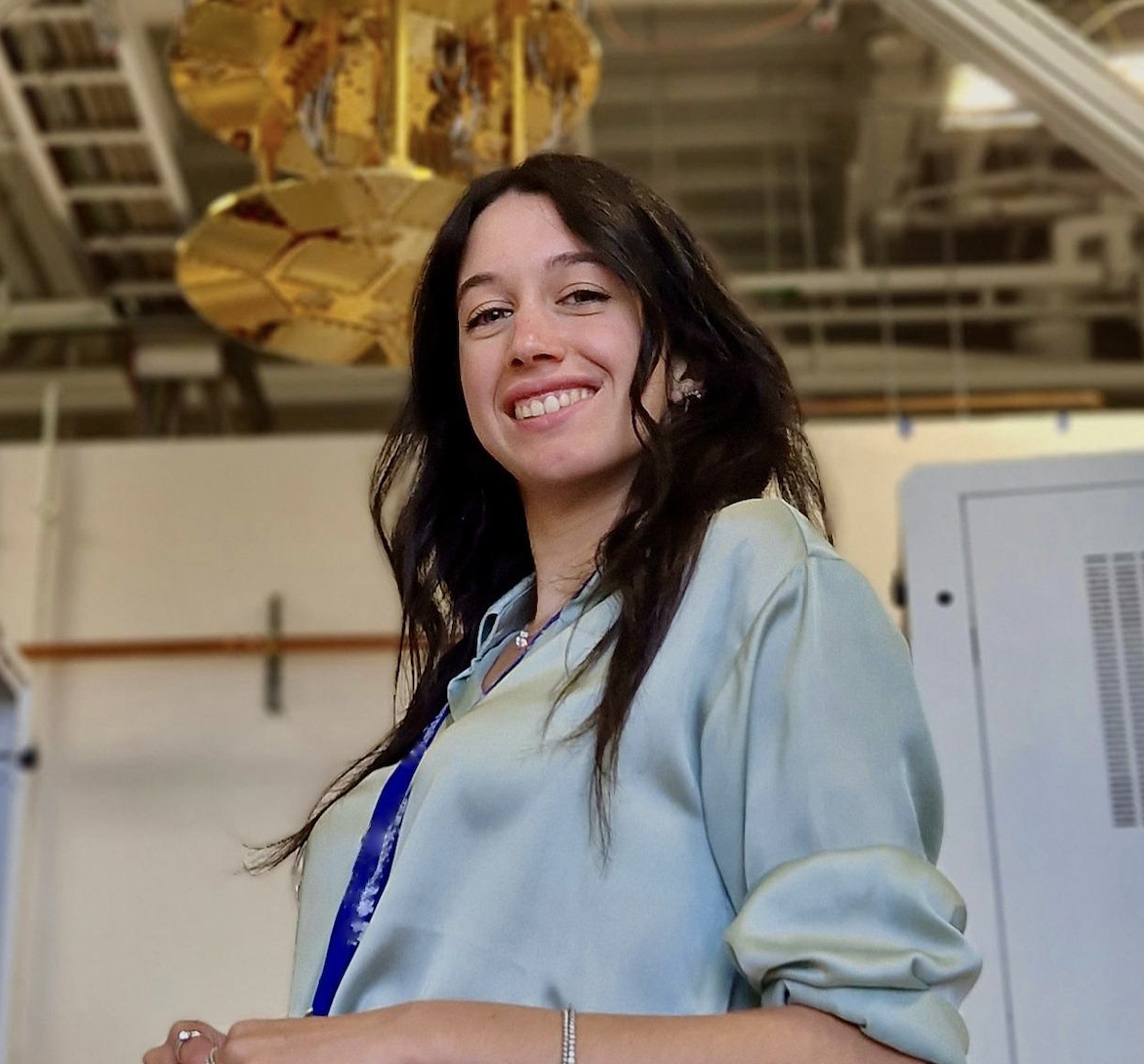}}]{Laura d'Avossa} (\textit{Graduate Student Member, IEEE}) Laura d’Avossa received the M.Sc degree in Electrical Engineering in 2022 (summa cum laude) from University of Naples Federico II (Italy). Since 2022 she is a member of  Quantum Internet Research Group (www.quantuminternet.it). In 2023, she was an Intern at Fermi National Accelerator Laboratory and in 2024 she served as a research assistant at Argonne National Laboratory, USA. In 2025, she was a visiting researcher at Northwestern University, USA.
In 2025 she received the Best Paper Award in the Quantum Networking \& Communications Track at IEEE Quantum Week (QCE).
Currently, she is a Ph.D. student within the Quantum Technologies doctoral program at the University of Naples Federico II. Her research focuses on quantum communications, particularly on experimental entanglement distribution and quantum transduction.
\end{IEEEbiography}

\begin{IEEEbiography}[{\includegraphics[width=1in,height=1.25in,clip,keepaspectratio]{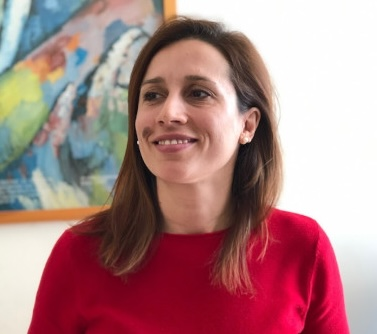}}]{Angela Sara Cacciapuoti} (\textit{Senior Member, IEEE})is currently a Full Professor of quantum communications and networks with the University of Naples Federico II, Naples, Italy and Co-Founder of Quantum Internet Research Group (www.quantuminternet.it).
She received the ERC Consolidator Grant “QNattyNet” (qnattynet.quantuminternet.it), which aims to lay the foundations of a truly quantum-native Internet. Her research focuses on the theoretical and architectural foundations of quantum networking, with contributions to quantum-network protocol design, resource management, and control architectures. She was the recipient of multiple honors recognizing pioneering work in the field. She was also recently recognized as a Featured Author on IEEE Xplore website. She has served as an IEEE ComSoc Distinguished  Lecturer, delivering invited talks worldwide on the Quantum Internet design. She is a member of Technical Committee on Signal Processing for Communications and Networking (SPCOM) within the IEEE Signal Processing Society. She serves as Area Editor of IEEE TRANSACTIONS ON COMMUNICATIONS and Senior Editor of IEEE JOURNAL ON SELECTED AREAS IN COMMUNICATIONS (JSAC) – Quantum Series. She is also on the editorial boards
of \textit{npj Quantum Information}, IEEE TRANSACTIONS ON QUANTUM ENGINEERING, and IEEE COMMUNICATIONS SURVEYS AND TUTORIALS. Previously, she
has held several IEEE leadership roles, including the Vice-Chair and Publicity Chair of WICE and Treasurer of IEEE Women in Engineering Affinity Group (Italy Section).
\end{IEEEbiography}

\bibliographystyle{IEEEtran}
\bibliography{bibliography.bib}
\end{document}